\documentclass [AMA,STIX1COL]{WileyNJD-v2}
\usepackage{moreverb}
\usepackage{moreverb,url}
\usepackage{amsmath}
\usepackage{kotex}
\usepackage{multirow}
\usepackage{graphicx}
\usepackage{amssymb}
\usepackage{relsize}
\usepackage{color}
\usepackage[LGR,T1]{fontenc}
\usepackage{bigstrut}
\usepackage{tikz}
\usepackage{verbatim}
\usepackage{float}
\usepackage{lineno,hyperref}
\usepackage{subfigure}
\usepackage{adjustbox}
\usepackage{caption}
\usepackage{setspace}
\usepackage{makecell}
\usepackage{lscape}
\usepackage[per -mode=symbol]{siunitx}
\usepackage{orcidlink}
\usepackage{hyperref}
\hypersetup{
colorlinks=true,
citecolor=blue,
linkcolor=blue,
urlcolor=blue
}

\newcommand\BibTeX{{\rmfamily B\kern-.05em \textsc{i\kern-.025em b}\kern-.08em
T\kern-.1667em\lower.7ex\hbox{E}\kern-.125emX}}

\articletype{RESEARCH ARTICLE}%

\received{<day> <Month>, <year>}
\revised{<day> <Month>, <year>}
\accepted{<day> <Month>, <year>}

\begin{document}

\title{Real-time response estimation of structural vibration with inverse force identification}

\author[1]{Seungin Oh}

\author[2]{Hanmin Lee}

\author[2]{Jai-Kyung Lee}

\author[3]{Hyungchul Yoon}

\author[1]{Jin-Gyun Kim*}

\authormark{SEUNGIN OH \textsc{et al}}

\address[1]{\orgdiv{Department of Mechanical Engineering (Integrated Engineering)}, \orgname{Kyung Hee University}, \orgaddress{\state{1732, Deogyeong-daero, Giheung-gu, Yongin-si, Gyeonggi-do 17104}, \country{Republic of Korea}}}

\address[2]{\orgdiv{Department of Smart Industrial Machine Technologies}, \orgname{Korea Institute of Machinery and Materials}, \orgaddress{\state{156 Gajeongbuk-ro, Yuseong-gu, Daejeon}, \country{Republic of Korea}}}

\address[3]{\orgdiv{School of Civil Engineering}, \orgname{Chungbuk National University}, \orgaddress{\state{1, Chungdae-ro, Seowon-gu, Cheongju-si, Chungcheongbuk-do 28644}, \country{Republic of Korea}}}

\corres{*Jin-Gyun Kim, Department of Mechanical Engineering (Integrated Engineering), Kyung Hee University, 1732, Deogyeong-daero, Giheung-gu, Yongin-si, Gyeonggi-do 17104, Korea. \email{jingyun.kim@khu.ac.kr}}

\abstract[Abstract]{This study aimed to develop a virtual sensing algorithm of structural vibration for the real-time identification of unmeasured information.
First, certain local point vibration responses (such as displacement and acceleration) are measured using physical sensors, and the data sets are extended using a numerical model to cover the unmeasured quantities through the entire spatial domain in the real-time computation process. 
A modified time integrator is then proposed to synchronize the physical sensors and the numerical model using inverse dynamics. 
In particular, an efficient inverse force identification method is derived using implicit time integration.
The second-order ordinary differential formulation and its projection-based reduced-order modeling is used to avoid two times larger degrees of freedom within the state space form. 
Then, the Tikhonov regularization noise-filtering algorithm is employed instead of Kalman filtering.
The performance of the proposed method is investigated on both numerical and experimental testbeds under sinusoidal and random excitation loading conditions.
In the experimental test, the algorithm is implemented on a single-board computer, including inverse force identification and unmeasured response prediction.
The results show that the virtual sensing algorithm can accurately identify unmeasured information, forces, and displacements throughout the vibration model in real time in a very limited computing environment.}

\keywords{Inverse force identification, Implicit method, Tikhonov regularization, Component mode synthesis, Virtual sensing, Digital twin}

\jnlcitation{\cname{%
\author{S. Oh}, 
\author{H. Lee}, 
\author{J. Lee}, 
\author{H. Yoon}, and 
\author{J.G. Kim}} (\cyear{2022}), 
\ctitle{Real-time response estimation of structural vibration with inverse force identification}, \cjournal{Arxiv}, \cvol{2022;00:0000}.}

\maketitle

\section{Introduction}\label{sec1}
Measuring structural vibration responses is essential for evaluating and predicting the condition of structures. Structural health monitoring (SHM) applications such as system identification and damage detection use vibration information to monitor structural conditions. Sensors, including accelerometers, LVDTs, and strain gauges, have been used for structural health monitoring purposes~\cite{doebling1996statistical, todd1999civil}. Furthermore, computer vision-based approaches have been replacing the method for measuring structural vibration for system identification purposes ~\cite{chen2015modal, feng2016vision, yoon2016target}. However, vibration measurement is still limited in certain cases owing to cost, inaccessibility, noise, and communication.
Virtual sensing, a real-time method for identifying and visualizing structural vibrations, is an attractive way to solve these issues.
This method can obtain the desired responses of a target system using only a few physical sensors.
Virtual sensing techniques have been proposed in various engineering fields, including avionics, traffic, automotive, building monitoring, chemical engineering, and industrial engineering ~\cite{Moreau2008, Kano2012,lei2020integration, Ahn2022}.
Conventional virtual sensing techniques have been used with simplified analytical (or numerical) models to compensate for the limited sensing data, known as model-based approaches.
Another virtual sensing technique relies on a data-driven (or empirical) approach.
This is based on reference datasets that employ ideal data, which enables damage and  fault sensor detection through comparison of the measurement and reference datasets ~\cite{Liu2009}.
In addition, noise reduction has also been studied using Kalman filtering~\cite{Lesniak2009}.
Data-driven approaches have been combined with machine learning methods such as support vector regression, deep neural networks (DNNs), and artificial neural networks (ANNs)~\cite{Atkinson1998}.
Machine-learning-based, data-driven approaches have been actively adopted for soft sensors~\cite{Yan2016, Jain2007}.
In terms of structural vibrations, Sun et al. recently proposed a data-driven virtual sensing technique for compensating  partially measured vibration data using a convolutional neural network (CNN)~\cite{Sun2017}.
However, machine learning-based methods require huge datasets to train their weights and bias, which means they are limited, except in certain cases when a sufficiently large dataset can be obtained.
Therefore, model-based and model-data combined with virtual sensors have been studied.
Lai et al. proposed a physics-informed system identification technique that  uses physical and neural network models to approximate the responses of ordinary differential equations~\cite{lai2021structural}.

For structural dynamics and vibration problems, model-based virtual sensing techniques~\cite{Risaliti2019,Kullaa2016,Oh2020} have been studied using various state identification algorithms such as Markov parameter-based and Kalman filter family methods.
%such as joint input-state estimation~\cite{Maes2015}, inverse force identification~\cite{liu2014explicit,lai2017explicit,Oh2020,lau1997, Zhu2000, Zhu2001, Law2001, Law2005}, Kalman filtering family methods~\cite{lourens2012augmented,Wei2022,Risaliti2019,chatzi2009unscented,yang2003constrained,azam2017experimental,shrivastava2020identification,naets2015stable, azam2015dual,erazo2017offline}, have been studied [너무 ref가 많음 -> Kalman + inverse force ID로 재편 필요].
Law et al.~\cite{lau1997} used an explicit time integration scheme to identify the moving forces in the time domain, and Zhu and Law implemented a method to identify the moving forces on multi-span bridges~\cite{Zhu2000,Zhu2001}. Law and Fang utilized dynamic programming techniques with state-space representation using the force identification method~\cite{Law2001}. Law et al. later proposed a system Markov parameter-based time-varying wind load identification method~\cite{Law2005}, and Kammer implemented a set of inverse Markov parameters for the inverse force identification technique~\cite{kammer1998input}.
Various researchers have also proposed Kalman filter-based identification methods.
Since the augmented Kalman filter (AKF) method proposed by Lourens et al.  ~\cite{lourens2012augmented}, various extensions have been studied.
 Azam et al. alleviated the numerical instabilities caused by unobservable responses by collocating dual Kalman filters to the identification process~\cite{azam2015dual}. Naets et al. added dummy measurement information to the input signal at the position level to stabilize the identification process~\cite{naets2015stable}.
However, most current methods are based on the state-space form and adopt explicit time integration methods.
These methods can provide high accuracy even in noisy infrastructure environments with relatively low dominant natural frequencies, such as multi-story buildings and bridges.
However, these methods also have disadvantages, such as a large discretization error with a low sampling frequency or a long sampling duration~\cite{liu2014explicit}. 
It becomes more severe in the case of stiff structures, which implies mid- and high-frequency dominant problems, such as mechanical components (rotor shaft, motor housing, drivetrain, etc.).
This can be alleviated by using a small time step size, but it incurs a huge computational cost for explicit methods with the state-space form.  
To address this problem, Liu et al. proposed an explicit-implicit formulation of inverse force identification based on the implicit Newmark–$\beta$ algorithm~\cite{liu2014explicit}, which is a hybrid technique that employs the advantages of the implicit method in the conventional explicit force identification process. 
The Tikhonov regularization method was then used to cancel the noise in second-order ordinary differential equations instead of the Kalman filter~\cite{tikhonov1963solution,tikhonov1995numerical}. 
Although the method can effectively identify the applied forces on such infrastructures, the method based on the Markov parameter may not be easy to apply to stiff structures owing to computational costs.

This study developed an implicit force and response identification algorithm for structural vibration, which implies a real-time virtual sensing system.
%The developed method is derived based on the implicit Newmark-$\beta$ integration scheme, and directly uses the second order differential equation form of the structural vibration. 
%A proper Reduced Order Modeling (ROM) technique is also implemented to the proposed force virtual sensing algorithm to alleviate the computational cost. 
%The measurement noise is controlled by the Tikhonov regularization method.
%In this work, the proposed method is also developed for experimental implementation by combining physical sensors, numerical model, and synchronization algorithm between hardware and software. 
%An example of the schematics is described in Figure~\ref{Schematics}.
%% ----------------------------------------
%%
%% Figure1. General schematics
%%
%% ----------------------------------------
%\begin{figure}[h] 
%    \centering
%    \includegraphics[width=0.9\textwidth]{Assets/schematics.eps}
%    \caption{\color{blue}An example of virtual sensing system. (그림에 displacement와 acceleartion 둘다 추가해야  함)}
%    \label{Schematics}
%\end{figure}
Predicting the unmeasured response data through the entire structural domain using only a few measured point datasets was conducted in the real-time online stage.  
A new modified time integrator was then developed in inverse dynamics to synchronize the numerical model with the real model. Measuring experimental displacement, velocity, and acceleration measurements is easier than an external loading measurement.
In the modified time integrator, unmeasured force identification for the future time step is essential for computing the unknown response.
An efficient and stable real-time inverse force identification process was derived by rearranging the conventional implicit time integration scheme. Then the Tikhonov regularization method~\cite{tikhonov1963solution,tikhonov1995numerical} was applied to the derived formulation for measurement error compensation.
Certain external forces computed by a few measured response datasets were used to compensate for the unloaded internal force in the proposed force identification scheme. 
The feasibility of the proposed scheme was examined using the well-known standard implicit Newmark-$\beta$ time integrator \cite{newmark1959method}.
We would like to emphasize that the other aim of this process can be applied to very limited computational environments, such as edge computing and single-board computers. 
The framework was first resolved into offline and online stages to achieve real-time virtual sensing with limited computational power. 
The offline stage covers finite element (FE) modeling and its reduced-order modeling (ROM), which was to pretreat highly demanded computational procedures.
The generated dynamic system matrices were then reduced using ROM techniques.
There are two main objectives of model reduction in this virtual sensing framework: (a) to meet requirement of low computation power applications and (b) to directly connect the virtual sensor with physical sensors.
To achieve these requirements, recent advanced ROM methods, which are extensions of conventional component mode synthesis (CMS) methods, have been considered~\cite{Craig1968, Kim2015(2), Kim2017}.
In this manner, the physical sensing positions are included as one of the master degrees of freedom (DOFs) in the reduced model without modal transformation. 
The master DOFs are then used as canals to reflect the measured response data to the virtual sensors in the online sensing stage.
%It should be also noted that the FE model generation and ROM processes, which are included in the time-consuming offline stage, are carried out in general computing environments.
%Only the final generated and reduced (mass, stiffness, and damping) matrices are embedded in a processor for \textit{real-time sensing} in the online stage.

The proposed algorithm was implemented and evaluated using both numerical and experimental testbeds.
For the numerical test, a motor housing structure with a complex geometry was used as the target structure to evaluate the practical utility of the system. 
The identification accuracy and computational efficiency were tested with various noise levels, and a comparison study between the proposed and widely used AKF methods was also performed. 
The system was experimentally implemented and evaluated using a simple cantilever beam problem.
The experimental testbed was constructed on a Raspberry Pi 3b+, widely used as a single-board computer.
A single-board computer has very limited computational resources; hence, the computational efficiency and stability of the proposed system under an extreme computational environment can be evaluated.
In both cases, the modified time integrating and applied force identification algorithms are embedded into the processor with reduced system matrices, and real-time online identification of unmeasured time-varying information is possible.

Section 2 describes the FE model generation and ROM procedures used in this study.
The proposed implicit inverse force identification technique is described in Section 3.
The numerical studies and experimental implementations are described in Sections 4 and 5, respectively. 
Finally, the conclusions are presented in Section 6.

\section{FE model generation and reduced order modeling}\label{section2}
%\section{{\color{red}Modeling: FE model generation and reduced-order modeling}}\label{section2}
Accuracy and efficiency are essential for real-time virtual sensors.
%Accuracy can be achieved by using a calibration process.
The accuracy of a virtual sensor depends on the reliability of the numerical model.
The target problem of this study addresses structural vibrations, and FE models are considered to develop a model-based virtual sensing system.
In addition, FE model reduction is important for real-time computation because an FE model with many DOFs is difficult to handle without experiencing delays in the online sensing process.
This section describes the theories of FE model generation and reduced-order modeling employed in this study.

The general equations of motion for the undamped structural dynamic system are expressed as follows:
%-----------------------------------------------------------------------------------------------------------
\begin{equation} \label{sysmat_ori}
    {\mathbf{M}\ddot{\mathbf{u}}+\mathbf{K}\mathbf{u}=\mathbf{f}},
\end{equation}
%-----------------------------------------------------------------------------------------------------------
where $\mathbf{M}$ and $\mathbf{K}$ are the mass and stiffness matrices, respectively, which are the system matrices. 
$\ddot{\mathbf{u}}$ and $\mathbf{u}$ are the acceleration and displacement vectors, respectively, and $\mathbf{f}$ is the applied force vector. 

%The accuracy of the virtual sensor depends on the validation level between the FE model and the real model.
%In this work, the FE model of the target structure is generated as the same geometry and mechanical properties that are directly measured from the real structure. Even the model is generated carefully with the directly measured quantities, inevitable modeling errors could occur according to various reasons including measurement and numerical errors. Hence, the generated FE model could be precisely pre-calibrated by using the various conventional FE model updating technologies to improve its reliability\cite{Friswell, Lim2016}.

%In the calibration process, the system matrices are updated by using modal data, such as the natural frequency, mode shape, FRF, modal damping coefficient, etc.

\begin{figure}[h] 
    \centering
    \includegraphics[width=0.6\textwidth]{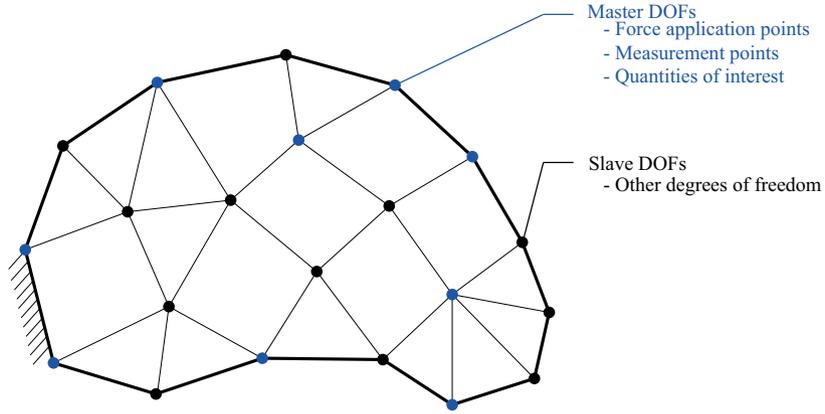}
    \caption{An example of selection of the master DOFs, blue dots: master DOFs, black dots: slave DOFs}
    \label{fig:Domains}
\end{figure}

%Alleviating the computational burden is a prerequisite to achieve a real-time virtual sensing system. 
The FE model generated with all DOFs in Eq. ~\eqref{sysmat_ori} is then approximated using ROM techniques for effective model updating and computational cost reduction.
In addition to computational efficiency, the direct connection between the reduced model and physical sensors is also important.
To synchronize numerical and real models in inverse dynamics, a general model reduction (GMR+) technique~\cite{Kim2017}, which is a hybrid method that combines dynamic condensation \cite{Guyan1965} and CMS~\cite{Craig1968, Kim2015(2), Go2020}, was considered.  
The DOFs of the full FE model were then divided into two subsystems: (a) master DOFs ($\mathbf{u}_m$) and (b) slave DOFs ($\mathbf{u}_s$).
All the DOFs with quantities of interest that need to be compared and identified in the physical domain should be included in the master DOFs, including loading and sensing points.  
All other DOFs were classified as slave DOFs ($\mathbf{u}_s$), as shown in Fig.~\ref{fig:Domains}. 

% ----------------------------------------
%
% Figure2. DOFs figure
%
% ----------------------------------------

The displacement vector in Eqs. ~\eqref{sysmat_ori} is expressed as
    \begin{equation} \label{disp_div}
        \mathbf{u} = \left[ \begin{array}{c} \mathbf{u}_{m} \\ \mathbf{u}_{s} \end{array} \right],
    \end{equation}
 and Eq.~\eqref{sysmat_ori} can be rewritten as follows:
    \begin{equation} \label{sysmat_div}
    \begin{aligned}
        \left[ \begin{array}{cc} {\mathbf{M}}_{mm} & {\mathbf{M}}_{ms} \\ {\mathbf{M}}_{sm} & {\mathbf{M}}_{ss} \end{array} \right]\left[ \begin{array}{c} \ddot{\mathbf{u}}_{m} \\ \ddot{\mathbf{u}}_{s} \end{array} \right]
        +
        \left[ \begin{array}{cc}  {\mathbf{K}}_{mm} & {\mathbf{K}}_{ms} \\ {\mathbf{K}}_{sm} & {\mathbf{K}}_{ss} \end{array} \right]\left[ \begin{array}{c} \mathbf{u}_{m} \\ \mathbf{u}_{s} \end{array} \right]=\left[ \begin{array}{c} \mathbf{f}_{m} \\ \mathbf{f}_{s} \end{array} \right],
        \end{aligned}
    \end{equation}
where the subscripts $m$ and $s$ are the master and slave DOFs, respectively. 
$N_m$ and $N_s$ are the number of master and slave DOFs, respectively ($N_m \ll N_s < N, N = N_m + N_s$). 
All master DOFs are preserved in the physical domain, but the slave DOFs are approximated using eigenbasis and constraint modes.
Subsequently, the displacement vector $\mathbf{u}$ in Eq. ~ \eqref{disp_div} in the GMR approach~\cite{Kim2017} can be written as:
    \begin{subequations} \label{ECB_disp}
        \begin{align}
        \mathbf{u} &= \left[ \begin{array}{c} \mathbf{u}_{m} \\ \mathbf{u}_{s} \end{array} \right] \approx   \hat{\mathbf{T}} \hat{\mathbf{u}}, \: \hat{\mathbf{T}} = \mathbf{T}_{0}+\mathbf{T}_r, \: \hat{\mathbf{u}} = \left[ \begin{array}{c} \mathbf{u}_{m} \\ {\mathbf{q}}_{s} \end{array} \right], \\
 \mathbf{T}_{0} &= \left[ \begin{array}{cc}  \mathbf{I} & \mathbf{0} \\ \mathbf{\Upsilon}  & \mathbf{\Psi}_d \end{array} \right], \: 
        \mathbf{T}_{r} = \left[ \begin{array}{cc} \mathbf{0} & \mathbf{0} \\  {\mathbf{F}_{rs}[ {\mathbf{M}}_{sm} +{\mathbf{M}}_{ss} \mathbf{\Upsilon}]} & \mathbf{0}  \end{array} \right]\hat{\mathbf{M}}_{0}^{-1}\hat{\mathbf{K}}_{0}, \:
\hat{\mathbf{M}}_{0} ={{\mathbf{T}_{0}}^T} \left[ \begin{array}{cc} {\mathbf{M}}_{mm} & {\mathbf{M}}_{ms} \\ {\mathbf{M}}_{sm} & {\mathbf{M}}_{ss} \end{array} \right] {\mathbf{T}_{0}}, \:
\hat{\mathbf{K}}_{0}={{\mathbf{T}_{0}}^T}\left[ \begin{array}{cc} {\mathbf{K}}_{mm} & {\mathbf{K}}_{ms} \\ {\mathbf{K}}_{sm} & {\mathbf{K}}_{ss} \end{array} \right]{\mathbf{T}_{0}}, \\
 \mathbf{\Psi}_{d}&=\left[ \begin{array}{ccc} \boldsymbol{\psi}_{1}&\dots&\boldsymbol{\psi}_{N_{d}}\end{array}\right], \: \mathbf{\Upsilon} = -{\mathbf{K}}^{-1}_{ss} {\mathbf{K}}_{sm}, \: \mathbf{F}_{rs} ={\mathbf{K}}_{ss}^{-1}-\mathbf{\Psi}_d\mathbf{\Gamma}_d^{-1}\mathbf{\Psi}_d^T, \: \mathbf{\Gamma}_{d}= \textit{diag} (\gamma_{1}, \dots, \gamma_{{N}_{d}}).
        \end{align}
    \end{subequations}

%    \begin{subequations} \label{u_s_approximation}
%        \begin{align}
%    \mathbf{u}_{s} \approx \mathbf{\Psi}_{d} {\mathbf{q}}_{s} + \mathbf{\Upsilon} \mathbf{u}_{m} + {\omega}^2 [ {\mathbf{M}}_{sm} +{\mathbf{M}}_{ss} \mathbf{\Upsilon}], \\ 
%     \mathbf{\Psi}_{d}=\left[ \begin{array}{ccc} \boldsymbol{\psi}_{1}&\dots&\boldsymbol{\psi}_{N_{d}}\end{array}\right], \: \mathbf{\Upsilon} = -{\mathbf{K}}^{-1}_{ss} {\mathbf{K}}_{sm},
%        \end{align}
%    \end{subequations}
where $\mathbf{F}_{rs}$ denotes the residual flexibility. 
$\mathbf{\Psi}_{d}$ and $\mathbf{\Upsilon}$ are the dominant eigenvector and constraint matrix, respectively.   
${\mathbf{q}}_{s}$ is the generalized coordinate vector of slave DOFs.
The $N_{d}$ eigenvectors of the slave DOFs can be calculated using the following eigenvalue problem:
 \begin{equation} \label{EVP_slave}
        {\mathbf{K}}_{ss}{\boldsymbol{\psi}}_i={\gamma}_i{\mathbf{M}}_{ss}\boldsymbol{\psi}_i~,~i=1,2,\dots,N_{d},
    \end{equation}
where $N_d$ denotes the number of selected dominant modes ($N_d\ll N_s$). 
$\boldsymbol{\psi}_{i}$ and $\gamma_{i}$ are the $i$th eigenvector and the eigenvalue, respectively.

Using $\hat{\mathbf{T}}$, the full equations in Eq. ~\eqref{sysmat_div} can be reduced to 
    \begin{subequations} \label{eq:reduction}
        \begin{align}
            &\hat{\mathbf{M}} \ddot{\hat{\mathbf{u}}}+\hat{\mathbf{K}} \hat{\mathbf{u}}=\hat{\mathbf{f}},  \\
            &\hat{\mathbf{M}}={\hat{\mathbf{T}}^T} \left[ \begin{array}{cc} {\mathbf{M}}_{mm} & {\mathbf{M}}_{ms} \\ {\mathbf{M}}_{sm} & {\mathbf{M}}_{ss} \end{array} \right] \hat{\mathbf{T}}, \: \hat{\mathbf{K}}={\hat{\mathbf{T}}^T}\left[ \begin{array}{cc} {\mathbf{K}}_{mm} & {\mathbf{K}}_{ms} \\ {\mathbf{K}}_{sm} & {\mathbf{K}}_{ss} \end{array} \right]\hat{\mathbf{T}}, \: \hat{\mathbf{f}}={\hat{\mathbf{T}}^T}\left[ \begin{array}{c} \mathbf{f}_{m}  \\ \mathbf{f}_{s} \end{array} \right]. \end{align}
    \end{subequations}
% and the eigenvalue problem of the reduced system is 
%    \begin{equation} \label{EVP_reduced}
%        \hat{\mathbf{K}}\hat{\boldsymbol{\phi}}_{i}=\hat{\lambda}_i\hat{\mathbf{M}}\hat{\boldsymbol{\phi}}_i~,~i=1,2,\dots,\hat{N},
%    \end{equation}
%    where $\hat{\lambda}_{i}$ and $\hat{\boldsymbol{\phi}}_{i}$ are the $i$th eigenvalue and eigenvector, respectively.   
 where the unknown vector $\hat{\mathbf{u}}$ includes the displacement vector of the master DOFs ($\mathbf{u}_{m}$) and modal coordinates of the slave DOFs (${\mathbf{q}}_{s}$). 
The number of DOFs in the reduced system is denoted by $\hat{N}$ ($\hat{N} \ll N, \hat{N} = N_{m}+N_d$). 
The number of DOFs in the reduced matrices was approximately 0.225 $\%$ of the total number of DOFs in the numerical study.
The size of the numerical model can be reduced to a reasonable number of DOFs to be computed in real-time through this reduction step.
The ROM process is similar to the Craig-Bampton (CB) method~\cite{Craig1968}, which is the most popular CMS method with primal assembly. 
However, the GMR formulation selects master DOFs similar to conventional dynamic condensation to apply the FE model-updating process appropriately.
Derivation details and comparison studies are described in \cite{Kim2017}. 

By adding viscous damping to Eqs. ~\eqref{eq:reduction}, the equations of motion of the damped structural dynamic system can be defined as
\begin{equation}\label{eq:reduced_sysmat_damped}
    \hat{\mathbf{M}} \ddot{\hat{\mathbf{u}}}+\hat{\mathbf{C}} \dot{\hat{\mathbf{u}}}+\hat{\mathbf{K}} \hat{\mathbf{u}}=\hat{\mathbf{f}}, \:\hat{\mathbf{C}}=a\hat{\mathbf{M}}+b\hat{\mathbf{K}},
\end{equation}
where $\hat{\mathbf{C}}$ denotes the viscous damping matrix. The mass and stiffness-proportional damping coefficients are $a$ and $b$, respectively. 

 The last step of the offline stage is to connect the reduced matrices directly, and the physical sensing process is a reordering of $\hat{\mathbf{u}}$ in Eq. ~\eqref{eq:reduced_sysmat_damped} as follows: 

    \begin{equation} \label{eq:displacement_reorganize_1}
        \hat{\mathbf{u}} =  \left[ \begin{array}{c} {\mathbf{u}}_{m} \\
                                                    {\mathbf{q}}_{s} \end{array} \right] =
                            \left[ \begin{array}{c} \left[ \begin{array}{c} {\mathbf{u}}_{m}^{m} \\
                                                    {\mathbf{u}}_{m}^{u} \\ \end{array}\right]\\
                                                    {\mathbf{q}}_{s} \end{array} \right].
    \end{equation}
    
This shows that the master displacement vector $\mathbf{u}_{m}$ is divided into two sub-vectors: the measured vectors $\mathbf{u}^{m}_{m}$ and an unmeasured vector $\mathbf{u}^{u}_{m}$. The superscripts $m$ and $u$ denote the measured and unmeasured quantities, respectively. 
It should be noted that among the master displacement vectors $\mathbf{u}_{m}$, only limited displacement information denoted by  $\mathbf{u}^{m}_{m}$ can be measured. 
Subsequently, the equations of motion in Eq. ~\eqref{eq:reduced_sysmat_damped} can be rewritten as
    \begin{equation}\label{eq:updated_divided_EOM_1}
            \begin{aligned}
        \left[ \begin{array}{ccc} \hat{\mathbf{M}}^{mm}_{mm} & \hat{\mathbf{M}}^{mu}_{mm} & \hat{\mathbf{M}}^{m}_{ms} \\
                        {[\hat{\mathbf{M}}^{mu}_{mm}]^{T}} & \hat{\mathbf{M}}^{uu}_{mm} &  \hat{\mathbf{M}}^{u}_{ms} \\
                        {[\hat{\mathbf{M}}^{m}_{ms}]^{T}} & {[\hat{\mathbf{M}}^{u}_{ms}]^{T}} & \hat{\mathbf{M}}_{ss} \end{array} \right]
        \left[ \begin{array}{c} \ddot{{\mathbf{u}}}_{m}^{m} \\
                                \ddot{{\mathbf{u}}}_{m}^{u} \\
                                \ddot{{\mathbf{q}}}_{s} \end{array} \right] 
        + 
        \left[ \begin{array}{ccc} \hat{\mathbf{C}}^{mm}_{mm} & \hat{\mathbf{C}}^{mu}_{mm} & \hat{\mathbf{C}}^{m}_{ms} \\
                                {[\hat{\mathbf{C}}^{mu}_{mm}]^{T}} & \hat{\mathbf{C}}^{uu}_{mm} &  \hat{\mathbf{C}}^{u}_{ms} \\
                                {[\hat{\mathbf{C}}^{m}_{ms}]^{T}} & {[\hat{\mathbf{C}}^{u}_{ms}]^{T}} & \hat{\mathbf{C}}_{ss} \end{array} \right]
        \left[ \begin{array}{c} \dot{{\mathbf{u}}}_{m}^{m} \\
                                \dot{{\mathbf{u}}}_{m}^{u} \\
                                \dot{{\mathbf{q}}}_{s} \end{array} \right] 
                                +
        \left[ \begin{array}{ccc} \hat{\mathbf{K}}^{mm}_{mm} & \hat{\mathbf{K}}^{mu}_{mm} & \hat{\mathbf{K}}^{m}_{ms} \\
                            {[\hat{\mathbf{K}}^{mu}_{mm}]^{T}} & \hat{\mathbf{K}}^{uu}_{mm} &  \hat{\mathbf{K}}^{u}_{ms} \\
                            {[\hat{\mathbf{K}}^{m}_{ms}]^{T}} & {[\hat{\mathbf{K}}^{u}_{ms}]^{T}} & \hat{\mathbf{K}}_{ss} \end{array} \right]
        \left[ \begin{array}{c} {\mathbf{u}}_{m}^{m} \\
                                {\mathbf{u}}_{m}^{u} \\
                                {\mathbf{q}}_{s} \end{array} \right]=
        \left[ \begin{array}{c} \hat{\mathbf{f}}^{m}_{m} \\
                                \hat{\mathbf{f}}^{u}_{m} \\
                                \hat{\mathbf{f}}_{s}\end{array} \right].
                                        \end{aligned}
    \end{equation}

The displacement vector in Eqs. ~\eqref{eq:updated_divided_EOM_1} should be reorganized into the measured and unmeasured subdomains. The physical DOFs, which were measured by the sensors, were redefined as a measured master displacement vector $\breve{\mathbf{u}}^{m}$. All other DOFs are combined as an unmeasured displacement vector $\breve{\mathbf{u}}^{u}$. The reorganized displacement vector is then expressed as

    \begin{equation} \label{eq:displacement_reorganize}
        \breve{\mathbf{u}} =  \left[ \begin{array}{c} \breve{\mathbf{u}}^{m} \\
                                                    \breve{\mathbf{u}}^{u} \end{array} \right], \: 
        \breve{\mathbf{u}}^{m} = {\mathbf{u}}_{m}^{m}, \: 
        \breve{\mathbf{u}}^{u} =  \left[ \begin{array}{c} {\mathbf{u}}_{m}^{u} \\
                                                    {\mathbf{q}}_{s} \end{array}\right],
    \end{equation}
    and the equations of motion in Eq. ~\eqref{eq:updated_divided_EOM_1} can be reorganized as follows:
     \begin{subequations}\label{eq:updated_divided_EOM}
    \begin{gather}
        \breve{\mathbf{M}}\ddot{\breve{\mathbf{u}}}+\breve{\mathbf{C}}\dot{\breve{\mathbf{u}}}+\breve{\mathbf{K}}\breve{\mathbf{u}}=\breve{\mathbf{f}},\\
        \breve{\mathbf{f}} =  \left[ \begin{array}{c} \breve{\mathbf{f}}^{m} \\
                                                    \breve{\mathbf{f}}^{u} \end{array} \right], \: 
        \breve{\mathbf{f}}^{m} = \hat{\mathbf{f}}_{m}^{m}, \: 
        \breve{\mathbf{f}}^{u} =  \left[ \begin{array}{c} \hat{\mathbf{f}}_{m}^{u} \\
                                                    \hat{\mathbf{f}}_{s} \end{array}\right],\\
        \breve{\mathbf{A}}=\left[ \begin{array}{cc} \breve{\mathbf{A}}^{m} & \breve{\mathbf{A}}^{c} \\ {[\breve{\mathbf{A}}^{c}]^{T}} & \breve{\mathbf{A}}^{u} \end{array} \right], ~
        \breve{\mathbf{A}}^{m}=\hat{\mathbf{A}}^{mm}_{mm},~
        \breve{\mathbf{A}}^{c}=\left[ \begin{array}{cc} \hat{\mathbf{A}}^{mu}_{mm} & \hat{\mathbf{A}}^{m}_{ms} \end{array} \right],~
        \breve{\mathbf{A}}^{u}=\left[ \begin{array}{cc}
        \hat{\mathbf{A}}^{uu}_{mm} &  \hat{\mathbf{A}}^{u}_{ms} \\
        {[\hat{\mathbf{A}}^{u}_{ms}]^{T}} & \hat{\mathbf{A}}_{ss} \end{array} \right],\\
        \{ \mathbf{M},\mathbf{C},\mathbf{K} \} \in \mathbf{A}.
        \end{gather}
    \end{subequations}
    
The modeling and reduction processes described in this section are the simplest examples that satisfy the generated FE model's minimum computational efficiency requirements and reliability. Hence, any updating and reduction methods or calibration processes can be replaced by more advanced or simple methodologies to improve the efficiency and accuracy of the numerical model.

% ----------------------------------------
%
% 3. Sensing algorithm
%
% ----------------------------------------

\section{Implicit methods with the force identification}\label{section3}
Synchronizing the numerical model with the target structure using a limited number of sensors is essential for achieving real-time virtual sensing of the unmeasured response.
This can be achieved using the time-integration method if the applied loading condition is measured. 
The conventional time integration method normally uses the applied force vector to compute the time-varying response of certain differential equations.
However, it is difficult to measure loading data in practice.
To overcome this issue, we propose a new time integrator that uses inverse dynamics with a force identification method.
The Tikhonov regularization method~\cite{tikhonov1963solution,tikhonov1995numerical} was applied to the derived formulation to compensate for the measurement error.
The applied force vector is identified using measured displacement, velocity, and acceleration responses.
The detailed applied force identification process and the modified time integrator are described in this section.

First, the general procedure of modified time integration is derived. 
In this work, the well-known Newmark-$\beta$ method is used among the various implicit time integration methods such as Newmark-$\beta$, trapezoidal, and Bathe's methods \cite{newmark1959method, Bathe2006}.
The force equilibrium equation at time ($t+\Delta t$) can be defined as
    
    \begin{equation} \label{eq:implicit_system}
        \breve{\mathbf{M}}{}^{t+\Delta t}\ddot{\breve{\mathbf{u}}}+\breve{\mathbf{C}}{}^{t+\Delta t}\dot{\breve{\mathbf{u}}}+\breve{\mathbf{K}}{}^{t+\Delta t}{\breve{\mathbf{u}}}={}^{t+\Delta t}\breve{\mathbf{f}},
    \end{equation}
    which is explicitly expressed as:
    \begin{equation} \label{eq:system_expansed}
    \begin{aligned}
        \left[ \begin{array}{cc} \breve{\mathbf{M}}^{m} & \breve{\mathbf{M}}^{c} \\ {[\breve{\mathbf{M}}^{c}]^{T}} & \breve{\mathbf{M}}^{u} \end{array} \right]
        \left[\begin{array}{c} {}^{t+\Delta t}\ddot{\breve{\mathbf{u}}}^{m} \\ {}^{t+\Delta t}\ddot{\breve{\mathbf{u}}}^{u} \end{array}\right]
        +
        \left[ \begin{array}{cc} \breve{\mathbf{C}}^{m} & \breve{\mathbf{C}}^{c} \\ {[\breve{\mathbf{C}}^{c}]^{T}} & \breve{\mathbf{C}}^{u} \end{array} \right]
        \left[\begin{array}{c} {}^{t+\Delta t}\dot{\breve{\mathbf{u}}}^{m} \\ {}^{t+\Delta t}\dot{\breve{\mathbf{u}}}^{u} \end{array}\right]
        +
        \left[ \begin{array}{cc} \breve{\mathbf{K}}^{m} & \breve{\mathbf{K}}^{c} \\ {[\breve{\mathbf{K}}^{c}]^{T}} & \breve{\mathbf{K}}^{u} \end{array} \right]
        \left[\begin{array}{c} {}^{t+\Delta t}\breve{\mathbf{u}}^{m} \\ {}^{t+\Delta t}\breve{\mathbf{u}}^{u} \end{array}\right]=
        \left[\begin{array}{c} {}^{t+\Delta t}\breve{\mathbf{f}}^{m} \\ {}^{t+\Delta t}\breve{\mathbf{f}}^{u} \end{array}\right].
    \end{aligned}
    \end{equation}

The velocity and acceleration vectors ($\dot{\breve{\mathbf{u}}}, \ddot{\breve{\mathbf{u}}}$) are defined by the numerical differentiation of the velocity and displacement vectors ($\dot{\breve{\mathbf{u}}}, {\breve{\mathbf{u}}}$) as

    \begin{subequations}\label{eq:response_t}
        \begin{align}
        {}^{t+\Delta t}\dot{\breve{\mathbf{u}}}&=\frac{\partial{}^{t+\Delta t}\breve{\mathbf{u}}}{\partial t}
        ={}^{t}\dot{\breve{\mathbf{u}}}+\Delta t(1-\delta){}^{t}\ddot{\breve{\mathbf{u}}}+\delta \Delta t{}^{t+\Delta t}\ddot{\breve{\mathbf{u}}},\\
        {}^{t+\Delta t}\ddot{\breve{\mathbf{u}}}&=\frac{\partial {}^{t+\Delta t}\dot{\breve{\mathbf{u}}}}{\partial t}
        =\frac{1}{\beta \Delta t^{2}}({}^{t+\Delta t}\breve{\mathbf{u}}-{}^{t}\breve{\mathbf{u}})-\frac{1}{\beta \Delta t}{}^{t}\dot{\breve{\mathbf{u}}}-(\frac{1}{2 \beta}-1){}^{t}\ddot{\breve{\mathbf{u}}}.
            \end{align}
        \end{subequations}

 Substituting Eq.~\eqref{eq:response_t} into Eq. ~\eqref{eq:system_expansed} and rearranging it, we obtain 

\begin{subequations} \label{eq:displacement_stand}
\begin{align}
        \tilde{\mathbf{K}} {}^{t+\Delta t}\breve{\mathbf{u}} &={}^{t+\Delta t}\tilde{\mathbf{f}}, \\
        \tilde{\mathbf{K}} &= \frac{1}{\beta \Delta t^{2}}\breve{\mathbf{M}}
        + \frac{\delta}{\beta \Delta t} \breve{\mathbf{C}}
        +\breve{\mathbf{K}},\\
        {}^{t+\Delta t}\tilde{\mathbf{f}} &=
        {}^{t+\Delta t}\breve{\mathbf{f}}+
        {}^{t+\Delta t}{\mathbf{r}}^{u},\\
        {}^{t+\Delta t}{\mathbf{r}}^{u}&=
        \breve{\mathbf{M}}
        (\frac{1}{\beta \Delta t^{2}}{}^{t}\breve{\mathbf{u}}
        + \frac{1}{\beta \Delta t}{}^{t}\dot{\breve{\mathbf{u}}}
        + (\frac{1}{2\beta}-1){}^{t}\ddot{\breve{\mathbf{u}}})
        +\breve{\mathbf{C}}
        (\frac{\delta}{\beta \Delta t}{}^{t}\breve{\mathbf{u}}
        + (\frac{\delta}{\beta}-1){}^{t}\dot{\breve{\mathbf{u}}}
        + (\frac{\delta\Delta t}{2\beta}-\Delta t){}^{t}\ddot{\breve{\mathbf{u}}}),
\end{align}    
\end{subequations}
    
    where $\tilde{\mathbf{K}}$ is an effective stiffness matrix and ${}^{t+\Delta t}\tilde{\mathbf{f}}$ is an effective force vector.

Here, ${}^{t+\Delta t}\dot{\breve{\mathbf{u}}}$ and ${}^{t+\Delta t}\ddot{\breve{\mathbf{u}}}$ are defined differently in the various implicit integration schemes. Thus, the formulation details in Eq. ~\eqref{eq:displacement_stand} can be replaced with various implicit integration schemes~\cite{newmark1959method, Bathe2006}.
In addition, we assume that the external force acts only on $\breve{\mathbf{u}}^{m}$, that is, ${}^{t+\Delta t} \breve{\mathbf{f}}^{u}=\mathbf{0}$.
Here, the internal force vectors ${}^{t+\Delta t}{\mathbf{r}}^{m}$ and ${}^{t+\Delta t}{\mathbf{r}}^{u}$ are known quantities using the data from the previous step.

The displacement vector of the next step (${}^{t+\Delta t}\breve{\mathbf{u}}$) can be calculated by solving Eq. ~\eqref{eq:displacement_stand}.
However, the external force vector, ${}^{t+\Delta t}\breve{\mathbf{f}}^{m}$ in Eq. ~\eqref{eq:displacement_stand} is unknown and difficult to measure in practice. Hence, it is identified by reprocessing directly measured response data using the inverse dynamics concept.
In this study, we consider it as the displacement data, which is ${}^{t+\Delta t}\breve{\mathbf{u}}^{m}$ in Eq. ~\eqref{eq:displacement_stand}.
To identify the applied forces, we first calculate the initial prediction of the displacement using internal forces with no external force assumption (${}^{t+\Delta t}\breve{\mathbf{f}}^{u}=\mathbf{0}$). 
Eq.~\eqref{eq:displacement_stand} can be expressed as:
    \begin{equation} \label{eq:displacement1_stand}
        \left[ \begin{array}{cc}\tilde{\mathbf{K}}^{m} & \tilde{\mathbf{K}}^{c} \\ \left[\tilde{\mathbf{K}}^{c}\right]^{T} & \tilde{\mathbf{K}}^{u} \end{array}\right]\left[\begin{array}{c} {}^{t+\Delta t}(\breve{\mathbf{u}}^{I})^{m} \\ {}^{t+\Delta t}(\breve{\mathbf{u}}^{I})^{u} \end{array}\right]=\left[\begin{array}{c} {}^{t+\Delta t}{\mathbf{r}}^{m} \\ {}^{t+\Delta t}{\mathbf{r}}^{u}\end{array}\right],
    \end{equation}
    where superscript $I$ denotes the quantities related to the initial prediction.
    The initial prediction of displacement(${}^{t+\Delta t}\breve{\mathbf{u}}^{I}$) can be calculated by solving Eq. ~\eqref{eq:displacement1_stand}, using information from the previous time step $t$. 
Subtracting Eq.~\eqref{eq:displacement1_stand}, from Eqs. ~\eqref{eq:displacement_stand} yields
    \begin{equation}\label{eq:difference1}
    \begin{aligned}
        \left[ \begin{array}{cc}\tilde{\mathbf{K}}^{m} & \tilde{\mathbf{K}}^{c} \\ \left[\tilde{\mathbf{K}}^{c}\right]^{T} & \tilde{\mathbf{K}}^{u} \end{array}\right]\left[\begin{array}{c} {}^{t+\Delta t}\Delta\breve{\mathbf{u}}^{m} \\ {}^{t+\Delta t}\Delta\breve{\mathbf{u}}^{u} \end{array}\right]=\left[\begin{array}{c} {}^{t+\Delta t}\breve{\mathbf{f}}^{m}\\ \mathbf{0}\end{array}\right],\\
        {}^{t+\Delta t}\Delta\breve{\mathbf{u}}={}^{t+\Delta t}\breve{\mathbf{u}} - {}^{t+\Delta t}\breve{\mathbf{u}}^{I}.
        \end{aligned}
        \end{equation}
    
    Then, from the second row of Eq. ~\eqref{eq:difference1}, $^{t+\Delta t}\Delta\breve{\mathbf{u}}^{u}$ is defined using only ${}^{t+\Delta t}\Delta\breve{\mathbf{u}}^{m}$ as follows:    
       \begin{equation}\label{eq:schure1-2}
        \begin{aligned}
            \left[ \tilde{\mathbf{K}}^{c} \right]^{T}{}^{t+\Delta t}\Delta\breve{\mathbf{u}}^{m}+\left[ \tilde{\mathbf{K}}^{u} \right]{}^{t+\Delta t}\Delta\breve{\mathbf{u}}^{u} = \mathbf{0}   \\
            \to {}^{t+\Delta t}\Delta\breve{\mathbf{u}}^{u}=-\left[ \tilde{\mathbf{K}}^{u} \right]^{-1}\left[ \tilde{\mathbf{K}}^{c} \right]^{T}{}^{t+\Delta t}\Delta\breve{\mathbf{u}}^{m}.
        \end{aligned}
    \end{equation}  
        
    Using Eq.~\eqref{eq:schure1-2} in the first row of Eq. ~\eqref{eq:difference1}, the unmeasured force vector ${}^{t+\Delta t} \breve{\mathbf{f}}^{m}$ can be indirectly computed as
     \begin{equation}\label{eq:estimated_force1}
        \begin{aligned}           
            {}^{t+\Delta t} \breve{\mathbf{f}}^{m} =  [\tilde{\mathbf{K}}^{m}-\tilde{\mathbf{K}}^{c}{[\tilde{\mathbf{K}}^{u}]^{-1}}[\tilde{\mathbf{K}}^{c}]^{T}] [{}^{t+\Delta t}\Delta\breve{\mathbf{u}}^{m}],
        \end{aligned}
    \end{equation}
       where ${}^{t+\Delta t}\breve{\mathbf{u}}^{m}$ is a known vector measured from the physical sensors of the target structure, and the applied force (${}^{t+\Delta t} \breve{\mathbf{f}}^{m}$) can be computed.

Owing to the ill-posedness of the inverse problem, the force directly identified through the proposed process can be highly unstable. Hence, to alleviate the instability of the identified solution, the Tikhonov regularization method~\cite{tikhonov1963solution,tikhonov1995numerical} was implemented.
To apply this method, the following damped least-squares formulation can be defined:
    \begin{equation}\label{eq:regularization}
    \begin{aligned}
        \mathbf{S}&=|| {}^{t+\Delta t}\breve{\mathbf{u}}^{m}_{e}-{}^{t+\Delta t}\breve{\mathbf{u}}^{m}_{n} ||^{2}+\alpha || {}^{t+\Delta t}\breve{\mathbf{f}}^{m} ||^{2},\\
        {}^{t+\Delta t}\breve{\mathbf{u}}^{m}_{n}&={}^{t+\Delta t}\Delta\breve{\mathbf{u}}^{m}_{n}+{}^{t+\Delta t}(\breve{\mathbf{u}}^{I})^{m}_{n},
        \end{aligned}
    \end{equation}
    where the subscripts $e$ and $n$ denote the experimentally and numerically obtained quantities, respectively, and $\alpha$ is the regularization parameter.
    It is essential to determine the appropriate value of $\alpha$ to identify a stable and accurate solution from the regularization process. In this study, the regularization parameter $\alpha$ was determined by the L-curve criterion~\cite{hansen1998rank}.
    Meanwhile, Eq.~\eqref{eq:estimated_force1} can be rewritten with respect to the measured displacement difference (${}^{t+\Delta t}\Delta\breve{\mathbf{u}}^{m}$) as follows:
\begin{equation}\label{eq:displacement_difference}
        \begin{aligned}           
             {}^{t+\Delta t}\Delta\breve{\mathbf{u}}^{m}=\mathbf{H}{}^{t+\Delta t} \breve{\mathbf{f}}^{m},~\mathbf{H}=[\tilde{\mathbf{K}}^{m}-\tilde{\mathbf{K}}^{c}{[\tilde{\mathbf{K}}^{u}]^{-1}}[\tilde{\mathbf{K}}^{c}]^{T}]^{-1}.
        \end{aligned}
    \end{equation}
    
    Using Eq.~\eqref{eq:displacement_difference}, the regularization equation in Eq. ~\eqref{eq:regularization} can be rewritten as
    \begin{equation}\label{eq:regularization_plugged}
    \begin{aligned}
        \mathbf{S}&=
        || {}^{t+\Delta t}\breve{\mathbf{u}}^{m}_{e}-(\mathbf{H}{}^{t+\Delta t}\breve{\mathbf{f}}^{m}+{}^{t+\Delta t}(\breve{\mathbf{u}}^{I})^{m}_{n}) ||^{2}
        +\alpha || {}^{t+\Delta t}\breve{\mathbf{f}}^{m} ||^{2}.
        \end{aligned}
    \end{equation}
    
To determine the optimal solution for the damped least-squares equation, Eq. ~\eqref{eq:regularization_plugged} can be differentiated by the force vector($\breve{\mathbf{f}}^{m}$) as: 
    \begin{equation}\label{eq:regularization_differentiated}
    \begin{aligned}
        \frac{\partial\mathbf{S}}{\partial\breve{\mathbf{f}}_{m}}&=-2\mathbf{H}^{T}({}^{t+\Delta t}\breve{\mathbf{u}}_{e}^{m}-(\mathbf{H}{}^{t+\Delta t} \breve{\mathbf{f}}^{m}+{}^{t+\Delta t}(\breve{\mathbf{u}}^{I})^{m}_{n}))
        +2\alpha({}^{t+\Delta t} \breve{\mathbf{f}}^{m})=0.
        \end{aligned}
    \end{equation}
    
    By rearranging Eq.~\eqref{eq:regularization_differentiated}, the regularized force vector ($\breve{\mathbf{f}}^{m}$) can be identified as follows:
    \begin{equation}\label{eq:forceid}
        {}^{t+\Delta t}\breve{\mathbf{f}}^{m}=(\mathbf{H}^{T}\mathbf{H}+\alpha\mathbf{I})^{-1}\mathbf{H}^{T}( {}^{t+\Delta t}\breve{\mathbf{u}}^{m}_{e}-{}^{t+\Delta t}(\breve{\mathbf{u}}^{I})^{m}_{n} ).
    \end{equation}

\begin{figure}[h] 
    \centering
    \includegraphics[width=0.7\textwidth]{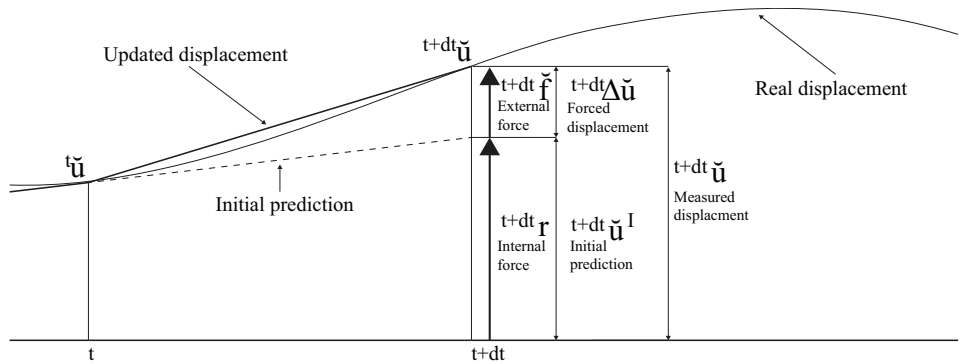}
    \caption{Integration schematics with a force identification}
    \label{fig:Integration_scheme}
\end{figure}

The entire effective force vector in Eqs. ~\eqref{eq:displacement_stand} is obtained by combining the identified force vector ${}^{t+\Delta t}\breve{\mathbf{f}}$ and internal force vector ${}^{t+\Delta t}\mathbf{r}$ as follows:
    \begin{equation}\label{eq:force_organize1}
        {}^{t+\Delta t}\tilde{\mathbf{f}} = {}^{t+\Delta t} \breve{\mathbf{f}} + {}^{t+\Delta t}\mathbf{r}=
        \left[\begin{array}{c} {}^{t+\Delta t}\breve{\mathbf{f}}^{m} \\ \mathbf{0}\end{array}\right]+\left[\begin{array}{c} {}^{t+\Delta t}\mathbf{r}^{m} \\ {}^{t+\Delta t}\mathbf{r}^{u}\end{array}\right],
    \end{equation}
Subsequently, $ {}^{t+\Delta t}\breve{\mathbf{u}}$ can be computed by solving Eq. ~\eqref{eq:displacement_stand}.
The velocity and acceleration vectors were computed using Eq. ~\eqref{eq:response_t}, and the process is iterated until the end. 
Finally, the unmeasured responses of the target structure were identified through the proposed time-integration process. Fig.~\ref{fig:Integration_scheme} shows a schematic representation of the proposed modified time-integration process, and the simplified procedure of the proposed algorithm is presented in Table. ~\ref{table:integration}.

This online process can synchronize a numerical model into a target structure in real-time.
In this study, the displacement data can be considered as measurement information in Eq. ~\eqref{eq:displacement_stand}; however, the velocity and acceleration data can be considered in engineering practice.
To derive the acceleration-based formulation by substituting the differentiation equation
Eq.\eqref{eq:response_t} into the displacement difference equation Eq~\eqref{eq:displacement_difference}, the force identification equation Eq.~\eqref{eq:forceid} can be redefined as:
    \begin{equation}
    \begin{aligned}
        {}^{t+\Delta t}\breve{\mathbf{f}}^{m}=&(({\beta^{2}\Delta t^{4}})\mathbf{H}^{T}\mathbf{H}
        +\alpha\mathbf{I})^{-1}\mathbf{H}^{T}({\beta\Delta t^{2}})( {}^{t+\Delta t}\ddot{\breve{\mathbf{u}}}^{m}_{e}-{}^{t+\Delta t}(\ddot{\breve{\mathbf{u}}}^{I})^{m}_{n} ).
        \end{aligned}
    \end{equation}

% ----------------------------------------
%
% Table1. Modified bathe time integration process.
%
% ----------------------------------------
\begin{table}[h]
    \centering
    \caption{Modified Newmark-$\beta$ time integration method with a force identification.}
    \begin{tabular}{l}
        \hline
        1. Initialization. \\
        \hline%\hline
        1-1. Initialize state vectors\\
        \hspace{3mm}${}^{0}\breve{\mathbf{u}}, {}^{0}\dot{\breve{\mathbf{u}}}, {}^{0}\ddot{\breve{\mathbf{u}}}$\\
        %\hline
        1-2. Calculate effective stiffness matrices\\
        \hspace{3mm}$\tilde{\mathbf{K}} = \frac{1}{\beta \Delta t^{2}}\breve{\mathbf{M}} + \frac{\delta}{\beta \Delta t}\breve{\mathbf{C}}+\breve{\mathbf{K}}$\\
        \hline%\hline
        2. Force identification.\\
        \hline%\hline
        2-1. Calculate internal force term\\
        \hspace{3mm}${}^{t+\Delta t}\mathbf{r}=\breve{\mathbf{M}}(\frac{1}{\beta \Delta t^{2}}{}^t\breve{\mathbf{u}} + \frac{1}{\beta \Delta t}{}^t\dot{\breve{\mathbf{u}}} + (\frac{1}{2\beta}-1){}^t\ddot{\breve{\mathbf{u}}})$\\
        $ +\breve{\mathbf{C}}(\frac{\delta}{\beta \Delta t}{}^t\breve{\mathbf{u}} + (\frac{\delta}{\beta}-1){}^t\dot{\breve{\mathbf{u}}} + (\frac{\delta \Delta t}{2\beta}-\Delta t){}^t\ddot{\breve{\mathbf{u}}})$\\
        2-2. Calculate initial prediction of displacement\\
        \hspace{3mm}${}^{t+\Delta t}\breve{\mathbf{u}}^I=[\tilde{\mathbf{K}}^{-1}]{}^{t+\Delta t}\mathbf{r}$\\
        %\hline
        2-3. Identify applied force\\
        \hspace{3mm}${}^{t+\Delta t}\breve{\mathbf{f}}^{m}=(\mathbf{H}^{T}\mathbf{H}+\alpha\mathbf{I})^{-1}\mathbf{H}^{T}( {}^{t+\Delta t}\breve{\mathbf{u}}^{m}_{e}-{}^{t+\Delta t}(\breve{\mathbf{u}}^{I})^{m}_{n} )$\\
        %\hline
        2-4. Calculate the effective force of the second sub-step\\
        \hspace{3mm}$^{t+\Delta t}\tilde{\mathbf{f}} ={}^{t+\Delta t}\breve{\mathbf{f}}+{}^{t+\Delta t}\mathbf{r},$\\
        \hspace{3mm}${}^{t+\Delta t}\breve{\mathbf{f}}=
        \left[\begin{array}{c}
        {}^{t+\Delta t}\breve{\mathbf{f}}^{m}\\
        \mathbf{0}\end{array} \right]$\\
        %\hline
        \hline%\hline
        3. Time integration.\\
        \hline%\hline
        3-1. Calculate state vectors of the second sub-step\\
        \hspace{3mm}$^{t+\Delta t}\breve{\mathbf{u}} =[\tilde{\mathbf{K}}^{-1}] ^{t+\Delta t}\tilde{\mathbf{f}}$\\
        \hspace{3mm}$^{t+\Delta t}\dot{\breve{\mathbf{u}}}={}^{t}\dot{\breve{\mathbf{u}}}+\Delta t(1-\delta){}^{t}\ddot{\breve{\mathbf{u}}}+\delta \Delta t{}^{t+\Delta t}\ddot{\breve{\mathbf{u}}}$\\
        \hspace{3mm}$^{t+\Delta t}\ddot{\breve{\mathbf{u}}}=\frac{1}{\beta \Delta t^{2}}({}^{t+\Delta t}\breve{\mathbf{u}}-{}^{t}\breve{\mathbf{u}})-\frac{1}{\beta \Delta t}{}^{t}\dot{\breve{\mathbf{u}}}-(\frac{1}{2 \beta}-1){}^{t}\ddot{\breve{\mathbf{u}}}$\\
        \hline
    \end{tabular}
    \label{table:integration}
\end{table}
% ----------------------------------------
%
% Figure4. Time integration procedure
%
% ----------------------------------------
%\begin{figure}
%    \centering
%    \includegraphics[width=0.45\textwidth]{Assets/Timeintegration_process.eps}
%    \caption{Online process of the modified time integration method with a force %identification.}    
%\label{fig:Timeintegration_procedure}
%\end{figure}

% ----------------------------------------
% 4. System modeling
% ----------------------------------------

\section{Numerical example}\label{section4}
The proposed method was evaluated using numerical and experimental tests.
The numerical tests are described in this section. 
The three aspects that were mainly considered in the numerical test are 1) the accuracy of the identified unmeasured responses, 2) stability under noisy conditions, and 3) computation efficiency.
In addition, a comparison test between the proposed method and the well-known AKF is performed~\cite{lourens2012augmented}.

\begin{figure}[h!]
    \centering
        \includegraphics[width=0.8\linewidth]{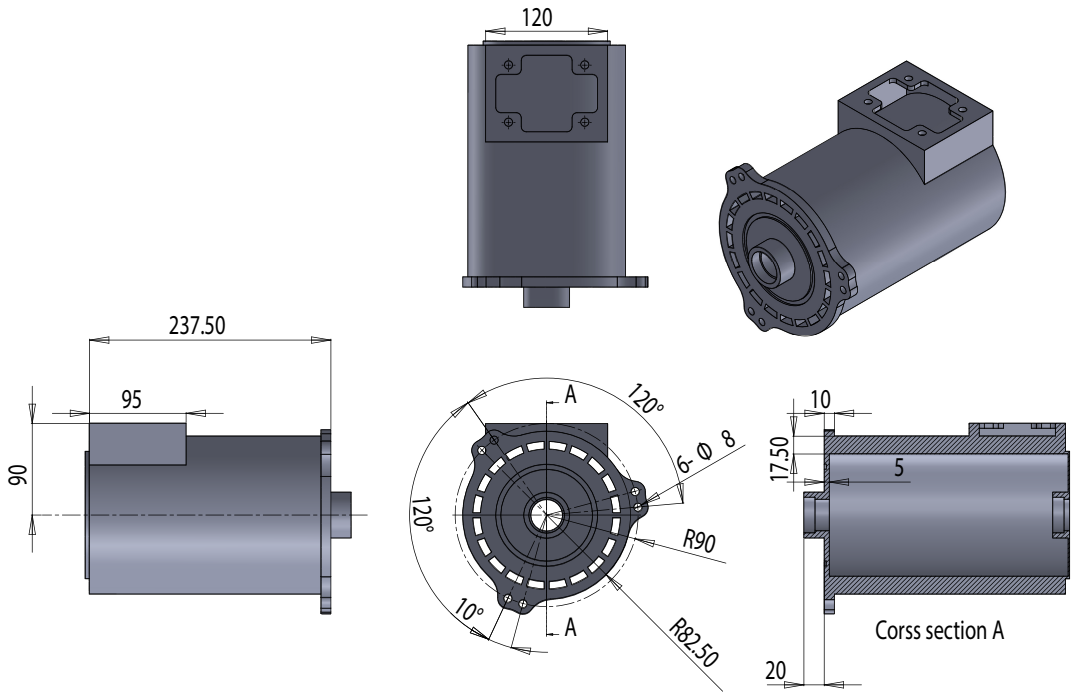}
        \caption{Geometry of numerical test model}
        \label{fig:numeric_model_geom}
\end{figure}

\begin{figure}[h!]
    \centering
        \includegraphics[width=0.8\linewidth]{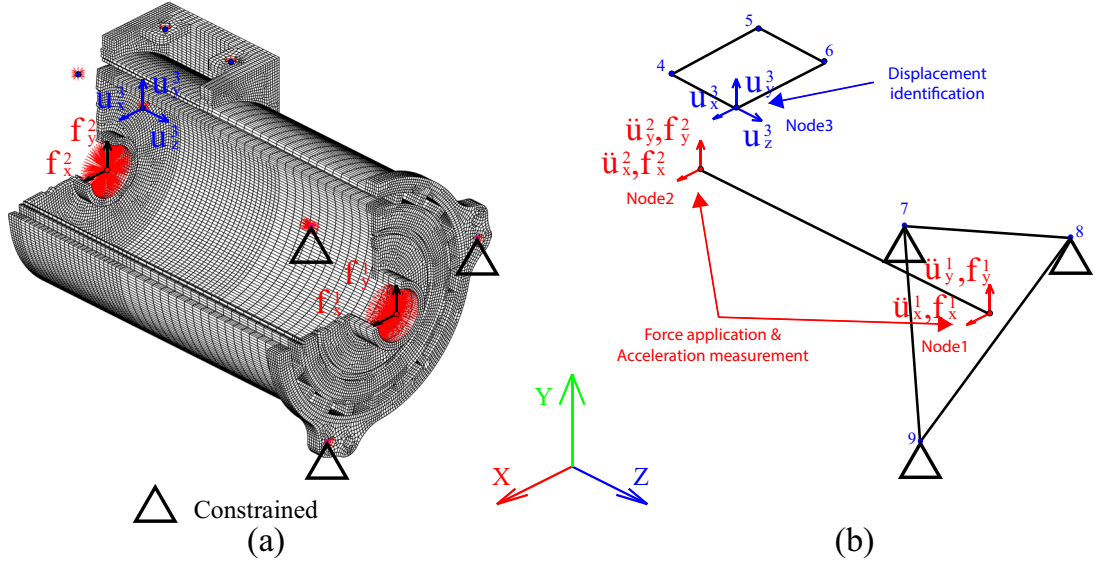}
        \caption{(a) Original full FE model and (b) the model after reduction}
        \label{fig:numeric_model}
\end{figure}

The target structure is the housing component of the motor system, and its detailed shape is shown in Fig.~\ref{fig:numeric_model_geom}.
 % ----------------------------------------
%
% Figure11. Results of the sinusoidal excitation cases
%
% ----------------------------------------
The FE mesh grid of the structure was generated using the 8-noded hexahedron solid elements, and the total counts of elements and nodes were 112,623 and 149,078, respectively. Young's modulus ($E$) and density ($\rho$) of the generated FE model were set to 210 $\si{\giga \pascal}$ and 8E-09 T/$\si{{\milli \meter}^{3}}$. Poisson’s ratio of the model was set to 0.3. 
Fig.~\ref{fig:numeric_model} shows the generated mesh grid and its reduced form.
The nodes around the holes are connected to the master nodes at the center of each hole using rigid body elements (RBE)~\cite{Ahn2020}. 
Constrained conditions were set on the three master nodes at the flange points of the structure, and external forces were applied to the RBE points ($Node 1$ and $Node 2$) of the bearing holes in the two orthogonal translational directions ($x$ and $y$).
% ----------------------------------------
%
% Figure11. Results of sinusoidal excitation cases
%
% ----------------------------------------

External forces were applied at the center of both bearing holes, and the loading profiles were set to the following functions:

\begin{subequations}\label{eq:force_comp}
    \begin{align}
    \mathbf{f}^{1}_{x}&=\left(1-exp^{\left(-\frac{t}{0.05}\right)}\right)\left(5500sin(2\pi \omega_{1}t^{2})+5000sin(2\pi \omega_{2} t)\right)\\
    \mathbf{f}^{1}_{y}&=\left(1-exp^{\left(-\frac{t}{0.05}\right)}\right)\left(5500cos(2\pi \omega_{1}t^{2})+5000sin(2\pi \omega_{2} t)\right)\\
    \mathbf{f}^{2}_{x}&=\left(1-exp^{\left(-\frac{t}{0.05}\right)}\right)\left(10000sin(2\pi \omega_{1}t^{2})+5400sin(2\pi \omega_{2} t)\right)\\
    \mathbf{f}^{2}_{y}&=\left(1-exp^{\left(-\frac{t}{0.05}\right)}\right)\left(10000sin(2\pi \omega_{1}t^{2})+5400sin(2\pi \omega_{2} t)\right)\\
    \omega_{1}&=1000~,\omega_{2}=200.
    \end{align}
\end{subequations}

To simulate the experimental signals of the target structure, the numerical solution of the defined problem was computed using the conventional Newmark-$\beta$ time integration method. 
%This is the pre-computation to simulate the experimentally measured signal, and that is separated from the numerical evaluation stage. 
The time increment value of the solution was set to 1E-07 $s$,  and the simulation duration was 0.0 to 0.1 $s$. Computed displacement and acceleration results of the simulation is collected from the \textit{`Measurement'} ($Node 1$ and $Node 2$) and \textit{`Identification reference'} ($Node 3$) points.

To consider the experimental signal noise, random Gaussian noise is added as
\begin{subequations}\label{eq:noise}
\begin{align}
    {}^{t}\bar{\mathbf{u}}={}^{t}{\mathbf{u}}+\sigma_{\mathbf{u}} \mathbf{N}_{noise},\\
    {}^{t}\ddot{\bar{\mathbf{u}}}={}^{t}\ddot{{\mathbf{u}}}+\sigma_{\ddot{\mathbf{u}}} \mathbf{N}_{noise},
    \end{align}
\end{subequations}
where $\sigma$ is the standard deviation of the induced noise and $\mathbf{N}_{noise}$ is the standard normal distribution function.
 Here, ${\mathbf{u}}$ and $\ddot{{\mathbf{u}}}$ were obtained from the numerical results without noise.
To quantify the levels of the added noise, the signal-to-noise ratio (SNR) could be defined as the following equation~\cite{Sherman2007}:
\begin{equation}
    \mathbf{SNR}_{\si{\deci\bel}}=20log_{10}\left[ \frac{\sqrt{\frac{\Sigma_{i=1}^{n_{signal}}(\bar{\mathbf{u}}_{i})^{2}}{n_{signal}}}}{\sqrt{\frac{\Sigma_{i=1}^{n_{signal}}((\bar{\mathbf{u}}_{i})^{2}-({\mathbf{u}}_{i})^{2})}{n_{signal}}}} \right].
\end{equation}

The generated FE model has 447,234 DOFs and is reduced using 54 master DOFs of RBE points and 47 normal modes, as described in Section 2. 
The master DOFs include 4 measurement DOFs ($\mathbf{u}^{1}_{x},\mathbf{u}^{1}_{y},\mathbf{u}^{2}_{x},\mathbf{u}^{2}_{y}$) and the other 50 DOFs at the RBE points as follows:
\begin{equation}
    \mathbf{u}_{m}=\left[\begin{array}{c} \mathbf{u}^{m}_{m} \\ \mathbf{u}^{u}_{m} \end{array}\right],~\mathbf{u}^{m}_{m}=\left[ \begin{array}{c}\mathbf{u}^{1}_{x}\\\mathbf{u}^{1}_{y}\\\mathbf{u}^{2}_{x}\\\mathbf{u}^{2}_{y}\end{array}\right],~
    \mathbf{u}^{u}_{m}=\left[ \begin{array}{c}\mathbf{u}^{1}_{z}\\\mathbf{u}^{1}_{\theta x} \\ \vdots \\ \mathbf{u}^{9}_{\theta x} \\ \mathbf{u}^{9}_{\theta y}\\ \mathbf{u}^{9}_{\theta z} \end{array} \right].
\end{equation}

\begin{figure}[h!] 
    \centering
    \includegraphics[width=0.5\linewidth]{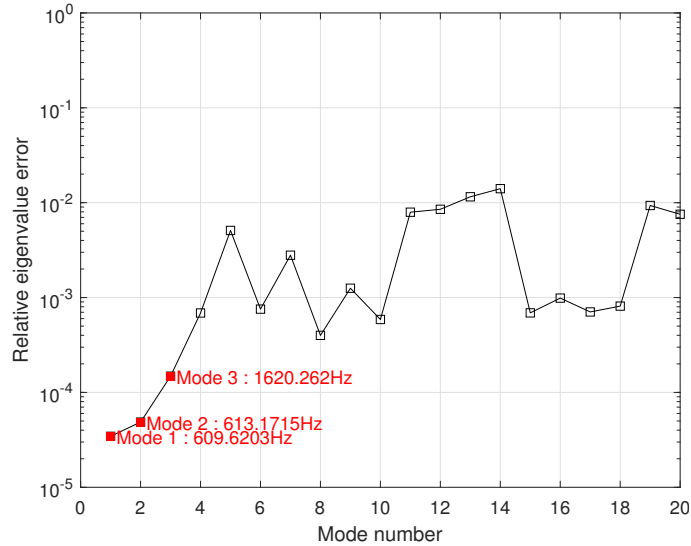}
    \caption{ Relative eigenvalue error of reduced model}
\label{fig:Reduced_error_numerical}
\end{figure}

Fig.~\ref{fig:Reduced_error_numerical} shows the relative eigenvalue errors between the full and reduced models.
The $i$th percentage eigenvalue error of the reduced system equations ($\hat{\epsilon}_{i}$) can be expressed as:
        \begin{equation}
        \hat{\epsilon}_{i}=\frac{{\lambda}_{i}-\hat{\lambda}_{i}}{{\lambda}_{i}}\times100.
        \end{equation}    
The results indicate that the reduced model can precisely preserve the dynamic properties of the original model.
% ----------------------------------------
%
% Figure9. Relative error of reduced model
%
% ----------------------------------------

A numerical test was performed using the acceleration signal as input. For the test, the \textit{measured} input signals $\ddot{\mathbf{u}}^{m}_{m}$ were collected from the pre-computed numerical solution as follows:  

\begin{equation}
    \ddot{{\mathbf{u}}}^{m}_{m}=\left[ \begin{array}{c}
    \ddot{\bar{\mathbf{u}}}^{1}_{x}\\
    \ddot{\bar{\mathbf{u}}}^{1}_{y}\\
    \ddot{\bar{\mathbf{u}}}^{2}_{x}\\
    \ddot{\bar{\mathbf{u}}}^{2}_{y}
    \end{array} \right].
\end{equation}

The desired force and displacement values ($\breve{\mathbf{f}}^{1,2}_{x,y},~\breve{\mathbf{u}}^{3}_{x,y,z}$) were then identified by performing the proposed identification process.
 In this study, the frequency domain error index (FDE) is used to quantify the accuracy of the identified results~\cite{dragovich2009fde,taher2021earthquake}:
\begin{equation}\label{eq:FDE}
        \mathbf{FDE}=\frac{\Sigma_{i=f_{0}}^{f_{max}}\sqrt{({R}_{ei}-{R}_{ci})^{2}+({I}_{ei}-{I}_{ci})^{2}}}{\Sigma_{i=f_{0}}^{f_{max}}(\sqrt{{R}_{ei}^{2}-{I}_{ei}^{2}}+\sqrt{{R}_{ci}^{2}-{I}_{ci}^{2}})},
        \end{equation}
        where $R$ and $I$ are the real and imaginary parts of the frequency spectrum data of the identified results, respectively, and the subscripts $e$ and $c$ denote the measured and identified quantities, respectively. $f_{0}$ and $f_{max}$ are the minimum and maximum frequencies of the obtained spectrum data, respectively.
The test conditions and their results are presented in the following sections. 

\subsection{Identification results}
\begin{figure}[h!]
    \centering
        \includegraphics[width=0.9\textwidth]{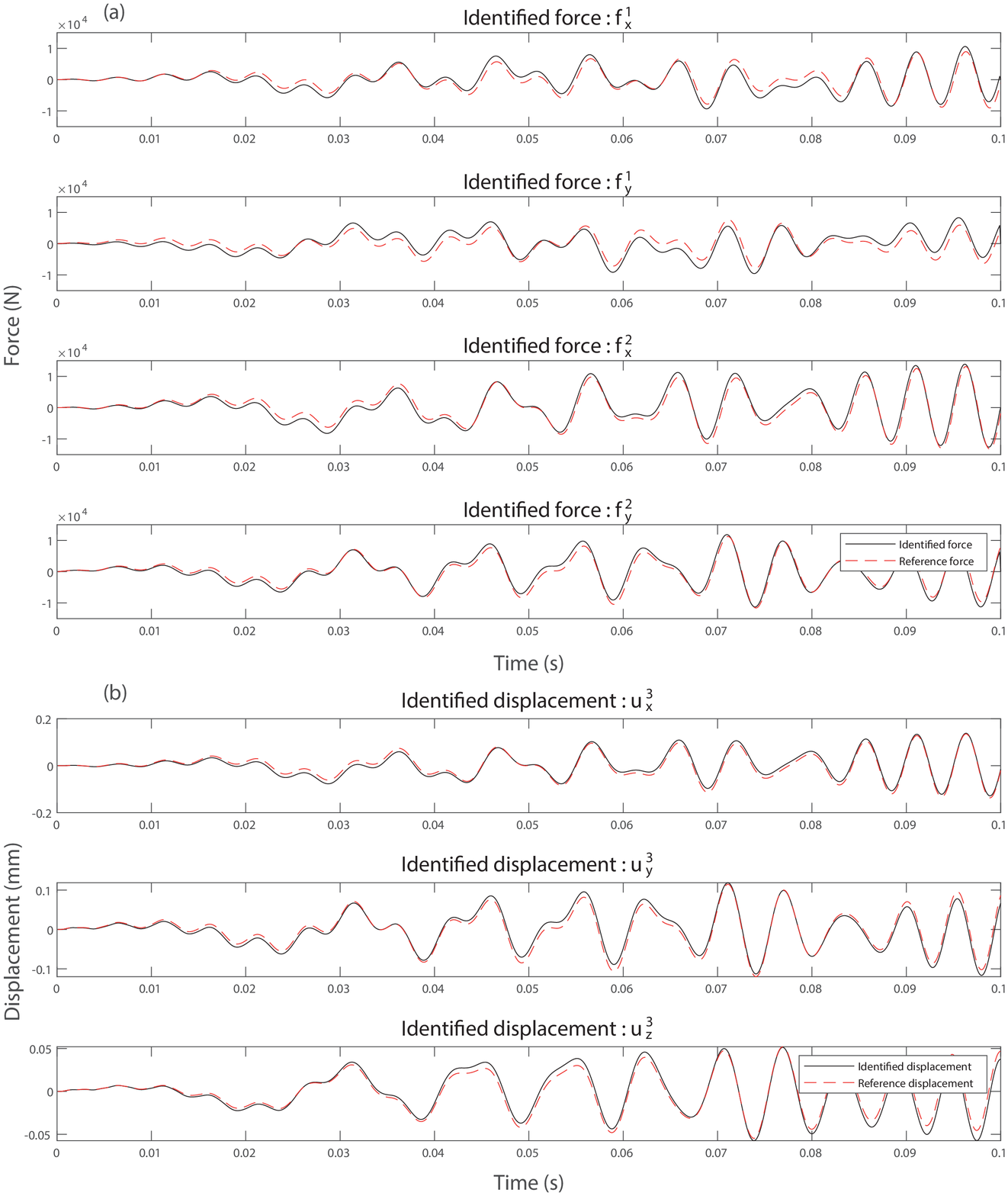}
        \caption{(a) Identified force results and (b) Identified displacement results using acceleration signals.}
        \label{fig:Numeric_results_acc}
\end{figure}

First, the unmeasured applied external forces were identified, and the unmeasured induced displacement signals were evaluated. 
The performance of the proposed method was also evaluated at various noise levels.
The standard deviation values of the acceleration error signals ($\sigma_{\ddot{\mathbf{u}}}$) were set to 0.5 $\%$ ($\sigma_{\ddot{\mathbf{u}}}$ = 0.005) of the standard deviation of the original signal; thus, the SNR of the applied error was set to 46 $\si{\deci\bel}$. 
The time increment value used for the identification test was set to 1E-04 $s$.
Fig.~\ref{fig:Numeric_results_acc}(a) presents the identified force results ($\breve{\mathbf{f}}^{1,2}_{x,y}$) and their reference values.
The results demonstrate that the proposed algorithm can effectively identify multiple applied forces.
The identified displacement signals ($\breve{\mathbf{u}}^{3}_{x,y,z}$) and their reference values are shown in Fig. ~\ref{fig:Numeric_results_acc}(b).
The results show that the identified displacements can also express the reference displacement signals well.
Table. ~\ref{table:FDE_acc} lists the FDE indices for the identified values.
Table.~\ref{table:etime_numeric} shows the entire calculation time and average computation time required for a single computation step in this algorithm.
The calculation time of the numerical test to identify 0.1 $s$ of the measured data is 0.0618 $s$, and the average computation time for a single step is 0.0618E-03 $s$. 
This demonstrates that the proposed method can identify the forces and responses faster than the required experimental sampling rate.

\begin{figure}[h!]
    \centering
    \includegraphics[width=0.9\textwidth]{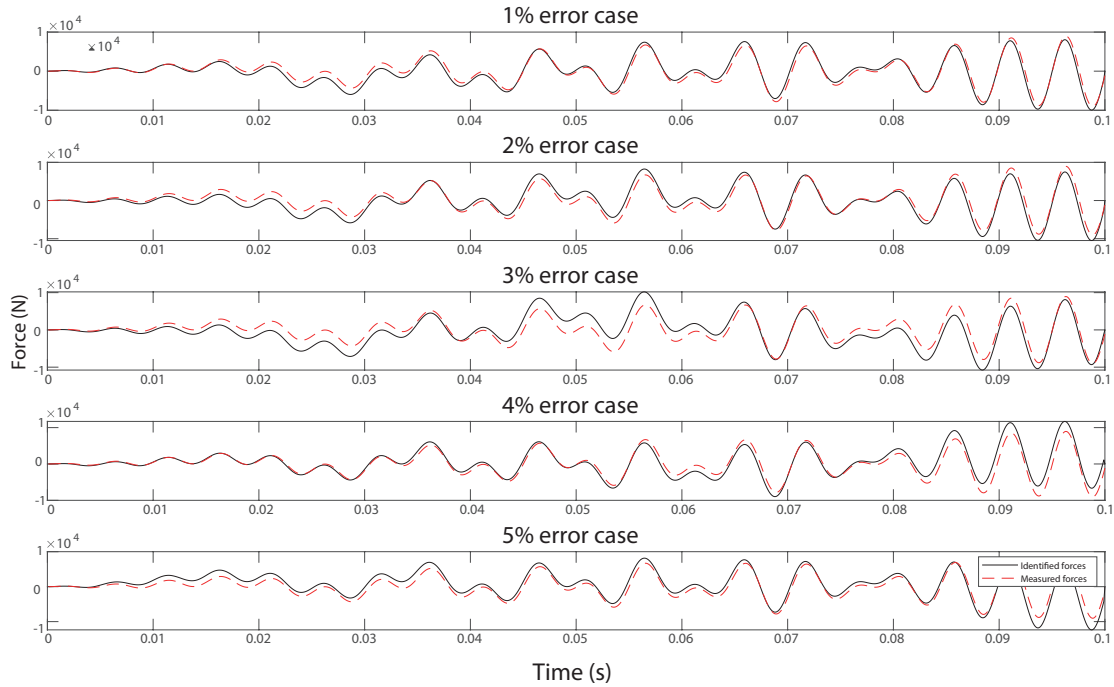}
    \caption{Identified force $\mathbf{f}_{x}^{1}$ results under various noise levels}
    \label{fig:noise_sensitivity_result}
\end{figure}

%\subsection{Robustness test results}
The performance of the proposed method at various noise levels was also evaluated. 
Acceleration signals were used as inputs, and five different noise level cases were considered.
The noise coefficients of the acceleration signals $\sigma_{\ddot{\mathbf{u}}}$ increase from 0.01  (1 $\%$, 40 $\si{\deci\bel}$) to 0.05 (5 $\%$, 26 $\si{\deci\bel}$) by the cases. First, the elements of the identified force values ($\mathbf{f}_{x}^{1}$) are collected and evaluated by comparing their analytical values in Eq. ~\ref{eq:force_comp}. 
The results show that the method can identify the applied forces, even under noisy conditions, as shown in Fig.~\ref{fig:noise_sensitivity_result}. 
The FDE indices of the system were also evaluated for the test cases. 
Table.~\ref{table:FDE_noise} presents the computed FDE values for each case.
The error-index values increase according to the increased error coefficients $\sigma_{\ddot{\mathbf{u}}}$, but all the error-index values are still below $0.2$.

% ----------------------------------------
%
% Figure11. Results of sinusoidal excitation cases
%
% ----------------------------------------

\begin{table}[h!]
    \centering
    \caption{FDE index values of acceleration based identification case}
 \begin{tabular}{c|ccccccc}
\hline
 Target value & $\mathbf{f}^{1}_{x}$ & $\mathbf{f}^{1}_{y}$ & $\mathbf{f}^{2}_{x}$ & $\mathbf{f}^{2}_{y}$ & $\mathbf{u}^{3}_{x}$ & $\mathbf{u}^{3}_{y}$ & $\mathbf{u}^{3}_{z}$\\
 \hline
FDE & 0.1460 & 0.1480 & 0.0865 & 0.0891 & 0.0363 & 0.0563 & 0.0595\\
\hline
\end{tabular}
    \label{table:FDE_acc}
\end{table}

%While performing the identification test, computational efficiency is also evaluated.
%The computational efficiency is one of the important criteria to evaluate the utility of the proposed system. Sufficiently efficient computation of the proposed identification process enables the real-time response estimation of the practical complex structure.
%The demanded sampling rate of the numerical testbed is set to $10,000 samples/sec$, and the total duration of the input signal is $0.1sec$. Calculation time to identify the responses is measured while performing the numerical test, and compared to the total duration of the simulation. {[\color{blue} sec, seconds, s 통일,그림도 마찬가지임]}

% ----------------------------------------
%
% Table 7. Computation time
%
% ----------------------------------------
        
\begin{table}[h!]
    \caption{Computation time of the virtual sensing system in the numerical test}
    \begin{center}
    \begin{tabular}{r r r}
    \hline
        Experimental time    & Calculation time & Single step time     \\ \hline
        0.1 s              &  0.0618 s      &  0.0618E-03 s (Avg.)        \\ \hline

    \end{tabular}\label{table:etime_numeric}
    \end{center}
\end{table}

\begin{table}[h!]
    \centering
    \caption{FDE index values under various noise levels}
 \begin{tabular}{c|ccccc}

\hline
 Noise levels & 1 $\%$ & 2 $\%$ & 3 $\%$ & 4 $\%$ & 5 $\%$ \\
 \hline
FDE & 0.0718 & 0.01046 & 0.1104 & 0.1397 & 0.1726\\
\hline
\end{tabular}
    \label{table:FDE_noise}
\end{table}

%\begin{figure}[h!]
%    \centering
%    \includegraphics[width=0.9\textwidth]{Assets/Noise_test/FDE_noise.eps}
%    \caption{FDE results under various noise levels}
%    \label{fig:noise_sensitivity_FDE}
%\end{figure}

\subsection{Comparative study}
A comparative study with other widely used force-identification methods was conducted to evaluate the relative performance of the proposed method.
The results of the proposed method are compared with those of the well-known AKF method\cite{lourens2012augmented} to compare their relative efficiency and robustness. 
The derivation of the conventional AKF method is explained in the original work\cite{lourens2012augmented}, and reduced equations of motion are implemented in the AKF method for equivalent comparison. 
The method proposed in this study was developed to identify the applied forces on relatively stiff components with high dominant eigenvalues. Therefore, the main consideration of the comparative study is the computational efficiency and the required time increment values for accurate identification. In the test, the time increment value of the proposed method was set to 1E-04 $s$, and the method could be stably implemented with this value owing to its implicit properties. In the case of the AKF method, the accuracy and stability of the identified results could be affected by the time increment values. Hence, various time increments (from 1E-07 to 1E-04 $s$) were tested for the AKF method to compare the efficiency and accuracy of the identified results. 1 $\%$ (40 $\si{\deci\bel}$) of Gaussian noise is applied to the input acceleration. 

\begin{figure}[h!]
    \centering
    \includegraphics[width=0.9\textwidth]{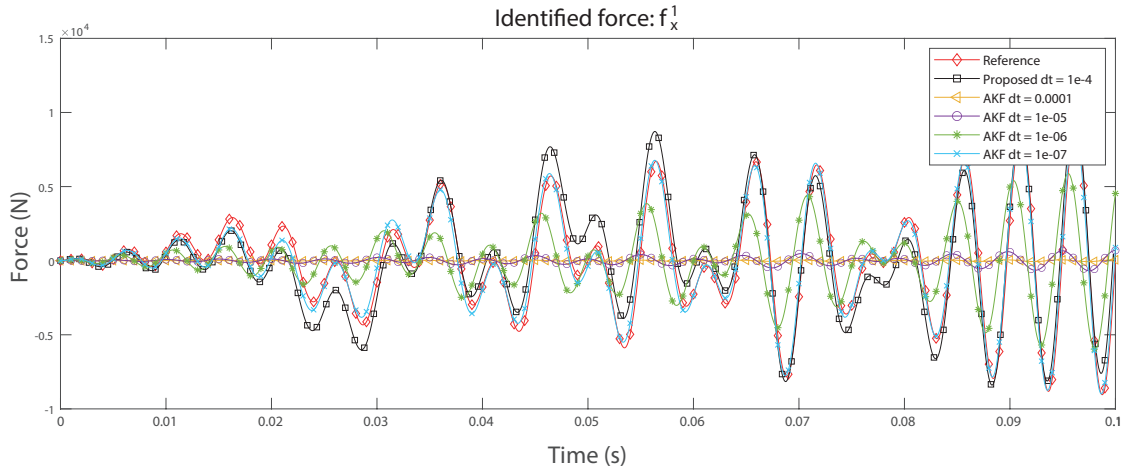}
    \caption{Comparative study results}
    \label{fig:Identifed_results_Comparative}
\end{figure}

The identified forces from each case were compared with the analytical values of Eq. ~\eqref{eq:force_comp}. Fig.~\ref{fig:Identifed_results_Comparative} shows the force element $\mathbf{f}_{x}^{1}$ of the results identified from the test cases.
The results show that the AKF method requires an approximately 1,000 times smaller time increment value to obtain results similar to the proposed method.
To compare the accuracy of the results, the FDE index values are computed using Eq. \eqref{eq:FDE}. Table.~\ref{table:Comparative_test_FDE} presents the computed FDE index values.
The method's efficiency was also evaluated by comparing the required computation time values of each case with the real duration of the reference signal (0.1 $s$). Table ~\ref{table:etime_FDE} shows the required computation time for the test cases.
Even with the same time increment sizes, the proposed algorithm is faster than the AKF method and approximately 16,644 times faster than the 1E-07 case. The results demonstrate that the proposed algorithm can be implemented more effectively on stiff structures with relatively high dominant eigenvalues.

\begin{table}[h!]
    \centering
    \caption{Frequency domain error results of comparative study cases}
 \begin{tabular}{c|c|c}
\hline
Used method &Time increments & Frequency domain error values\\
\hline
   Proposed method &1e-04 & 0.1110\\
\hline
& 1e-04 & 0.9917\\
AKF& 5e-05 & 0.9248\\
& 1e-06 & 0.6370\\
& 1e-07 & 0.1386\\
\hline
\end{tabular}
    \label{table:Comparative_test_FDE}
\end{table}

\begin{table}[h!]
    \centering
    \caption{Required computation time for identification}
 \begin{tabular}{c|c|c}
\hline
Used method &Time increments &Required time for computation ($s$)\\
\hline
   Proposed method &1e-04 & 0.0678\\
\hline
& 1e-04 & 0.4737\\
AKF& 5e-05 & 4.6566\\
& 1e-06 & 48.3128\\
& 1e-07 & 511.8409\\
\hline
\end{tabular}
    \label{table:etime_FDE}
\end{table}

\section{Experimental example}\label{section4}

\begin{figure}[h!]
    \centering
        \includegraphics[width=0.8 \linewidth]{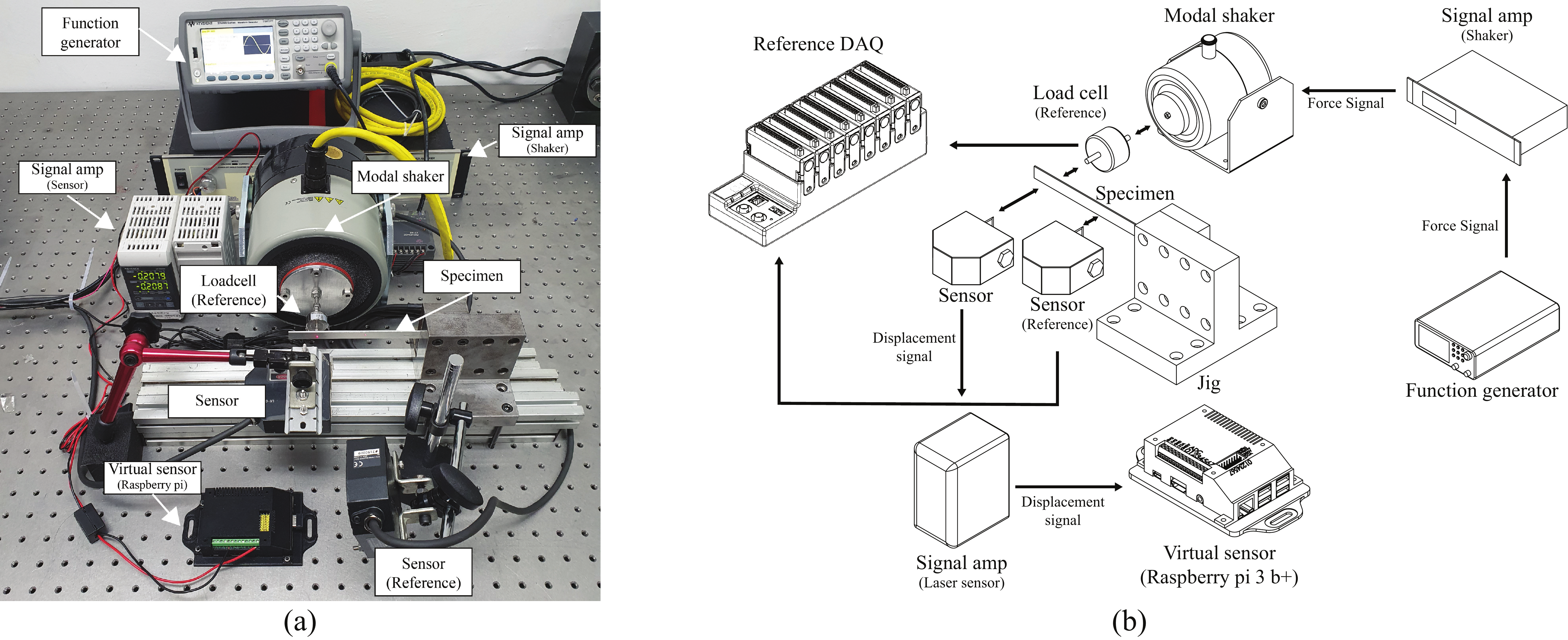}
        \caption{(a) Experiment setup, (b) Schematic diagram of experimental setup}
        \label{fig:EXP_setup}
\end{figure}

The proposed algorithm was also implemented and evaluated on a real test rig, as shown in Fig.~\ref{fig:EXP_setup}.
The test rig consisted of an example target structure, a virtual sensor processing unit, sensors to measure the response of the target structure and additional related instruments for the validation process.

\subsection{Test rig design}

\begin{figure}[h!] 
        \centering
        \includegraphics[width=0.8\linewidth]{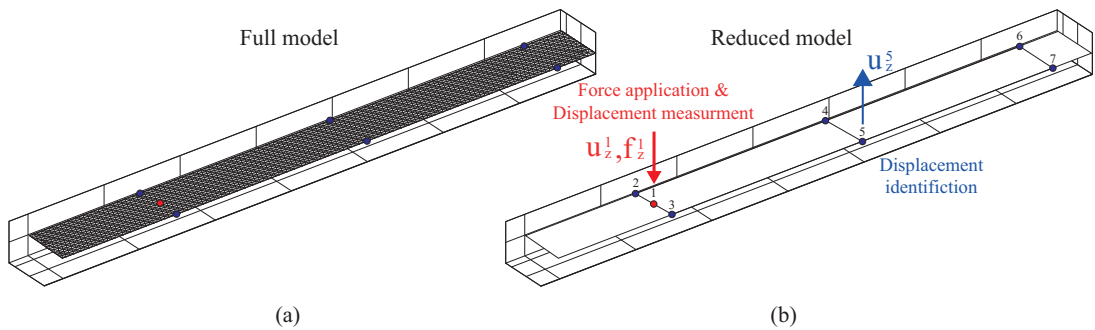}
        \caption{\label{fig:Reduced_geometry} (a) Original full FE model and (b) the model after reduction}
    \end{figure}

The target structure was a simple cantilever beam made of a thin aluminum plate, and the structure was fixed on a stiff steel jig at the end of the structure to apply a fixed boundary condition. The beam was 170 $\si{\milli \meter}$, the width of the structure was 13 $\si{\milli \meter}$, and the thickness was 1.2 $\si{\milli \meter}$.
The FE model is generated using the same geometry, and mass and stiffness matrices are then generated using the Mindlin-Reissner plate theory with triangular elements~\cite{bathe2006finite}.
The elastic modulus $E$ is initialized as 69 $\si{\giga \pascal}$ (updated later in the calibration stage), and Poisson's ratio ($\nu$) is 0.3. 
The density $\rho$ of the mass matrix $\mathbf{M}$ was computed using the measured weight of the target structure.
The generated FE model was then reduced by performing the ROM process to overcome the limitations of computational resources.
 First, the master DOFs ($\mathbf{u}_{m}$) are chosen, which include a single DOF for the measurement point ($\mathbf{u}^{m}_{m}$) and other response identification points ($\mathbf{u}^{u}_{m}$). The chosen master DOFs can be explicitly expressed as:
\begin{equation}
    \mathbf{u}_{m}=\left[\begin{array}{c} \mathbf{u}^{m}_{m} \\ \mathbf{u}^{u}_{m} \end{array}\right],~\mathbf{u}^{m}_{m}=\mathbf{u}^{1}_{z},~\mathbf{u}^{u}_{m}=\left[ \begin{array}{c}\mathbf{u}^{1}_{\theta x}\\\mathbf{u}^{1}_{\theta y} \\ \vdots \\ \mathbf{u}^{7}_{z} \\ \mathbf{u}^{7}_{\theta x}\\ \mathbf{u}^{7}_{\theta y} \end{array} \right].
\end{equation}
Thus, 21 DOFs were chosen as the master DOFs, including one measured point and 20 unmeasured DOFs. Fig.~\ref{fig:Reduced_geometry} shows the full and reduced FE models of the target structure, and the chosen unmeasured and measured master DOFs are marked as blue and red dots, respectively.

% ----------------------------------------
%
% Figure5. Experimental setup
%
% ----------------------------------------

% ----------------------------------------
%
% Figure8. Reduced geometry
%
% ----------------------------------------
    
\begin{figure}[h!] 
    \centering
    \includegraphics[width=0.5\linewidth]{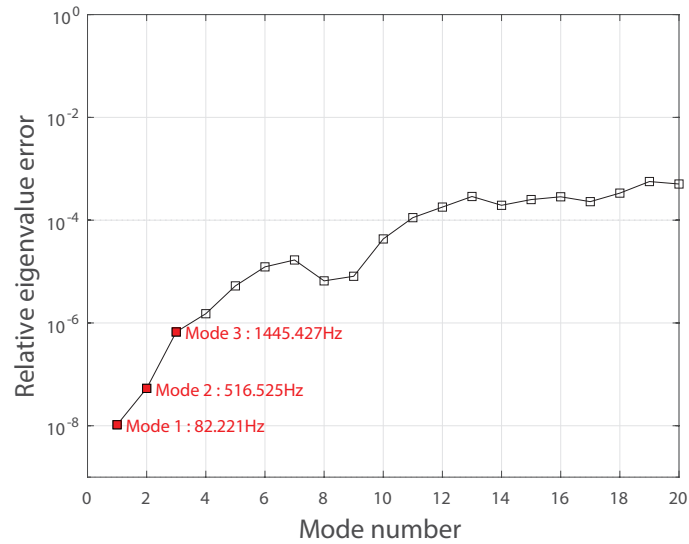}
    \caption{\label{fig:Reduced_error} Relative eigenvalue error of reduced model}
\end{figure}

  The 50 lowest eigenmodes were used to construct the reduced matrices.
  Using Eq.~\eqref{eq:reduction}, the FE model is reduced to the GMR+ domain.
    The number of DOFs in the model was then reduced from 5,487 to 71, which included 21 master DOFs and 50 modal DOFs.  Fig.~\ref{fig:Reduced_error} shows the relative eigenvalue errors between the full and reduced models.
These results demonstrate that the reduced model can precisely preserve the accuracy of the original model.
The reduced FE model is then reduced using the GMR+ method and implanted into an online virtual sensor processing unit.
% ----------------------------------------
%
% Figure9. Relative error of reduced model
%
% ----------------------------------------

The virtual sensor processing unit consists of instruments, including the main processor (Raspberry Pi 3 b+) and a signal acquisition module (ADS1256 analog-to-digital converter). A single-board computer (SBC) is the main processor, providing the computational resources needed to perform the virtual sensing process.
Raspberry Pi, a widely used, low-cost SBC, was chosen to meet the cost efficiency and stability demands of the developed system in a limited computation environment. The detailed specifications of the Raspberry Pi 3 b+ model are listed in Table~\ref{table:RPI_spec}.
The modified time integration process for the online identification stage presented in Section~\ref{section3} using custom-made software was implemented in the main processor using the C+ language. The target calculation capability was 1,000 $Samples/s$, and custom software was developed to ensure real-time virtual sensing calculations with the required sample rate.

% ----------------------------------------
%
% Table 2. Hardware specifications of Raspberry pi 3 b+
%
% ----------------------------------------
\begin{table}[h!]
    \caption{Hardware specifications of Raspberry Pi 3 b+}
    \begin{center}
    \begin{tabular}{c c c}
    \hline
        CPU & RAM & OS \\ \hline
        1.4$\si{\giga\Hz}$ 4 cores (ARM Cortex-A53) & 1GB & Linux   \\ \hline
    \end{tabular}\label{table:RPI_spec}
    \end{center}
\end{table}

Real-time displacement data were obtained using the signal acquisition module to synchronize the generated FE model with the target structure.
A laser sensor (LK-G30, KEYENCE) was used to measure the displacement response at a point in the target structure (110 $\si{\milli \meter}$ from the fixed edge). The measured displacement was used as the source displacement ($\breve{\mathbf{u}}^{m}$) for the identification process. The displacement information from this sensor is the source displacement data used to perform the suggested identification process and is the only information needed for practical use. Simultaneously, an additional laser displacement sensor (LK-G150, KEYENCE) was used to measure the real displacement response in the middle of the structure (60 $\si{\milli \meter}$ away from the fixed edge).
This additional measurement was only used as a reference to evaluate the accuracy of the identified unmeasured displacement ($\breve{\mathbf{u}}^{u}$). It was not necessary for the implementation of the proposed algorithm. 

\begin{figure}[h!] 
    \centering
    \includegraphics[width=0.6\linewidth]{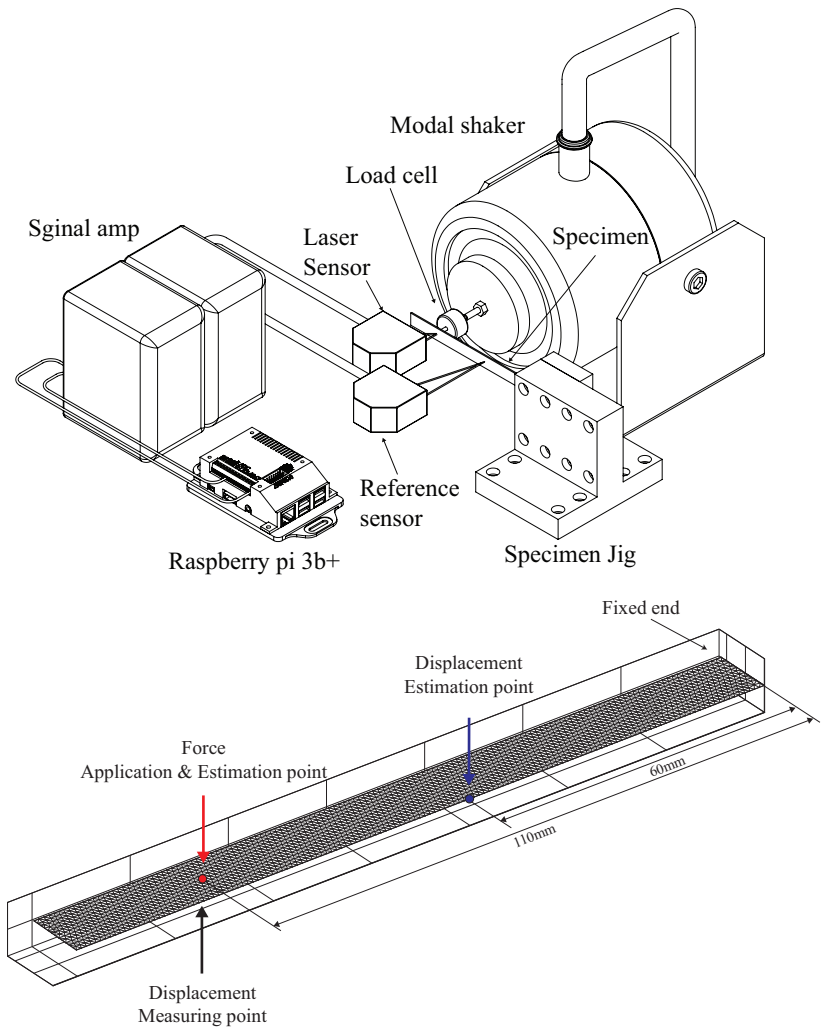}
    \caption{Experimental schematics}
    \label{fig:EXP_Schem}
\end{figure}

The target structure was excited with a modal shaker (ET-139, LabWorks) in the direction normal to the plate surface, with an intended loading profile at the same position as the displacement measuring point. 
The excitation signal was generated using a function generator (33500 B, Keysight) and amplified using a signal amplifier (PA-138, LabWorks). 
The applied force ($\breve{\mathbf{f}}^{m}$) was measured by using a load cell (UMM 5 $\si{\kilo\gram}$f, DACELL).
The measured force was also used as a reference signal to evaluate the accuracy of the identified applied force in the proposed virtual sensing. 
The reference displacement and force data were measured using an independent signal acquisition system (cDAQ, National Instruments). 
Fig.~\ref{fig:EXP_Schem} shows the detailed structure of the entire test rig and the test specimen. The positions of the force application and displacement identification points are marked by red and blue dots, respectively.

% --------------------
%
% Figure6. EXP_Schematics
%
% ----------------------------------------

% ----------------------------------------
%
% 5. Validation experiment cases
%
% ----------------------------------------
\subsection{Experimental result}\label{subsection_evaluation}
In this section, we evaluate the performance of the developed virtual sensing system using the designed test rig.
Similar to the numerical test, the accuracy and efficiency of the constructed virtual sensing system were evaluated by comparing the identified results with directly measured quantities at the identification point on the real structure.
This excitation test included sinusoidal and random noise excitation tests to check the accuracy and stability under noisy and periodic loading conditions.
The real-time identification performance was tested by measuring the calculation time of the system, and the results were compared with the actual duration of the experiment. Details of the test cases are presented in the following sections.

A simple sinusoidal loading condition was used as the first step in evaluating the proposed virtual sensing system. The lowest excitation loading frequency was 5 $\si{\Hz}$, and the frequency was doubled for each case up to 40 $\si{\Hz}$.

\begin{figure}[h!]
    \centering
        \includegraphics[width=0.9\textwidth]{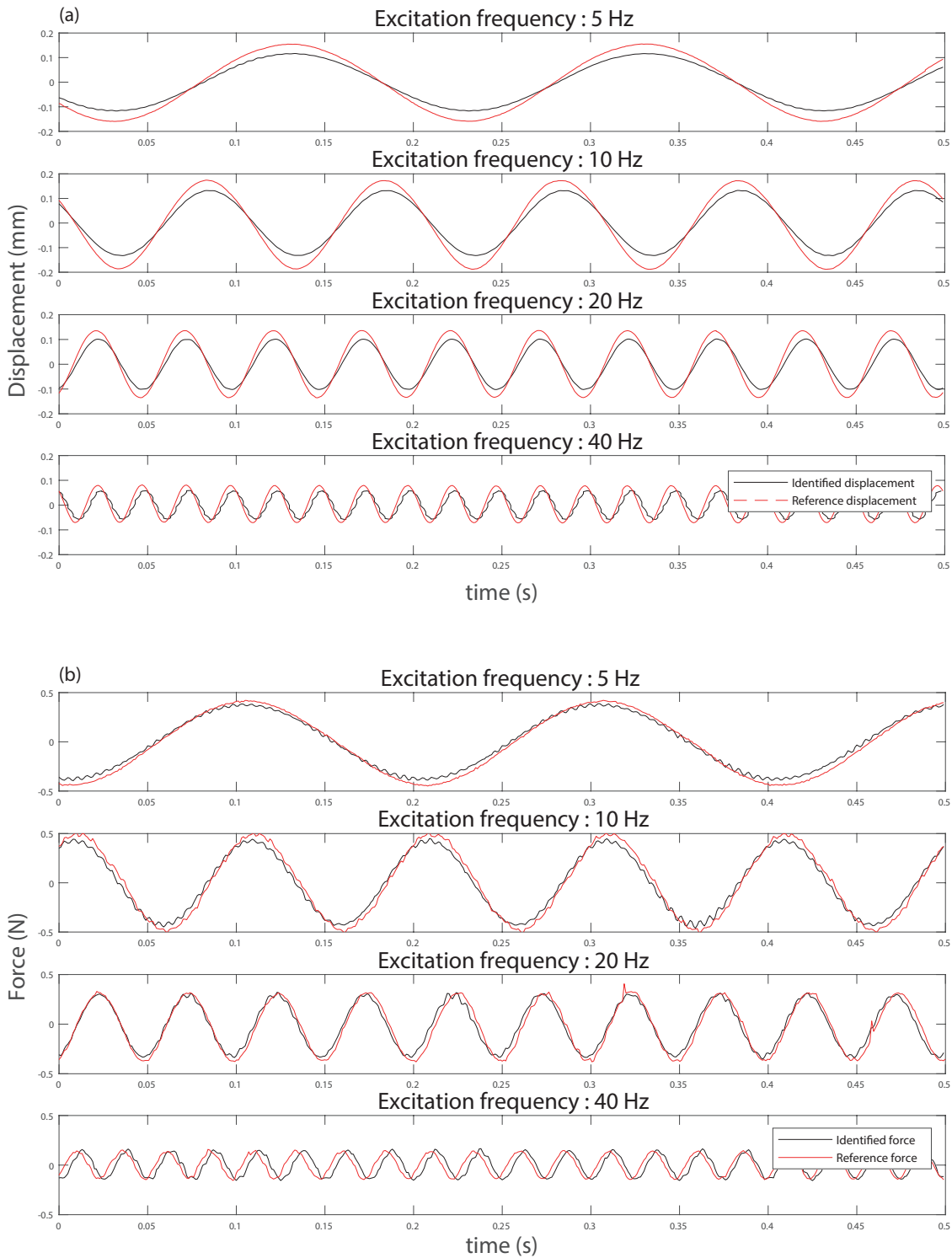}
        \caption{Results of sinusoidal excitation cases. (a) Displacement result plot of sinusoidal loading cases, (b) Force result plot of sinusoidal loading cases.}
        \label{fig:Result_sin}
\end{figure}

The identified and measured displacements are shown in Fig. ~\ref{fig:Result_sin}(a). This indicates that the identified displacement results agree with the real reference displacements in the three different excitation loading cases.
The identified force results for the sinusoidal loading cases are shown in Fig.~\ref{fig:Result_sin}(b). Some fluctuations were observed in the identified results, but the trends in the values were similar to those of  the measured force values. 
These results demonstrate that the developed system can stably and accurately identify unmeasured responses under sinusoidal excitation conditions.
Table ~\ref{table:FDE_exp_sinu} presents the FDE index values of the identified quantities under sinusoidal excitation.

\begin{table}[h!]
    \centering
    \caption{FDE index values of sinusoidal cases}
 \begin{tabular}{c|cccc}

\hline
 Excitation frequency & 5 $\si{\Hz}$ & 10 $\si{\Hz}$ & 20 $\si{\Hz}$ & 40 $\si{\Hz}$\\
 \hline
Identified force & 0.0848 & 0.1749 & 0.1716 & 0.2240 \\
Identified displacement & 0.1547 & 0.1501 & 0.1480 & 0.1510\\
\hline
\end{tabular}
    \label{table:FDE_exp_sinu}
\end{table}

Various excitation loading conditions can cause numerical instability in virtual sensing systems. Many mechanical systems undergo unstable loading conditions, including high-frequency noise. 
The virtual sensor system was tested under random loading conditions to validate the robustness of the developed system in a noisy environment.
The random excitation cases were performed with eight different random loading conditions, and the frequency range started from 5 $\si{\Hz}$ and was doubled for each case up to 160 $\si{\Hz}$.

\begin{figure}[h!]
    \centering
        \includegraphics[width=0.9\textwidth]{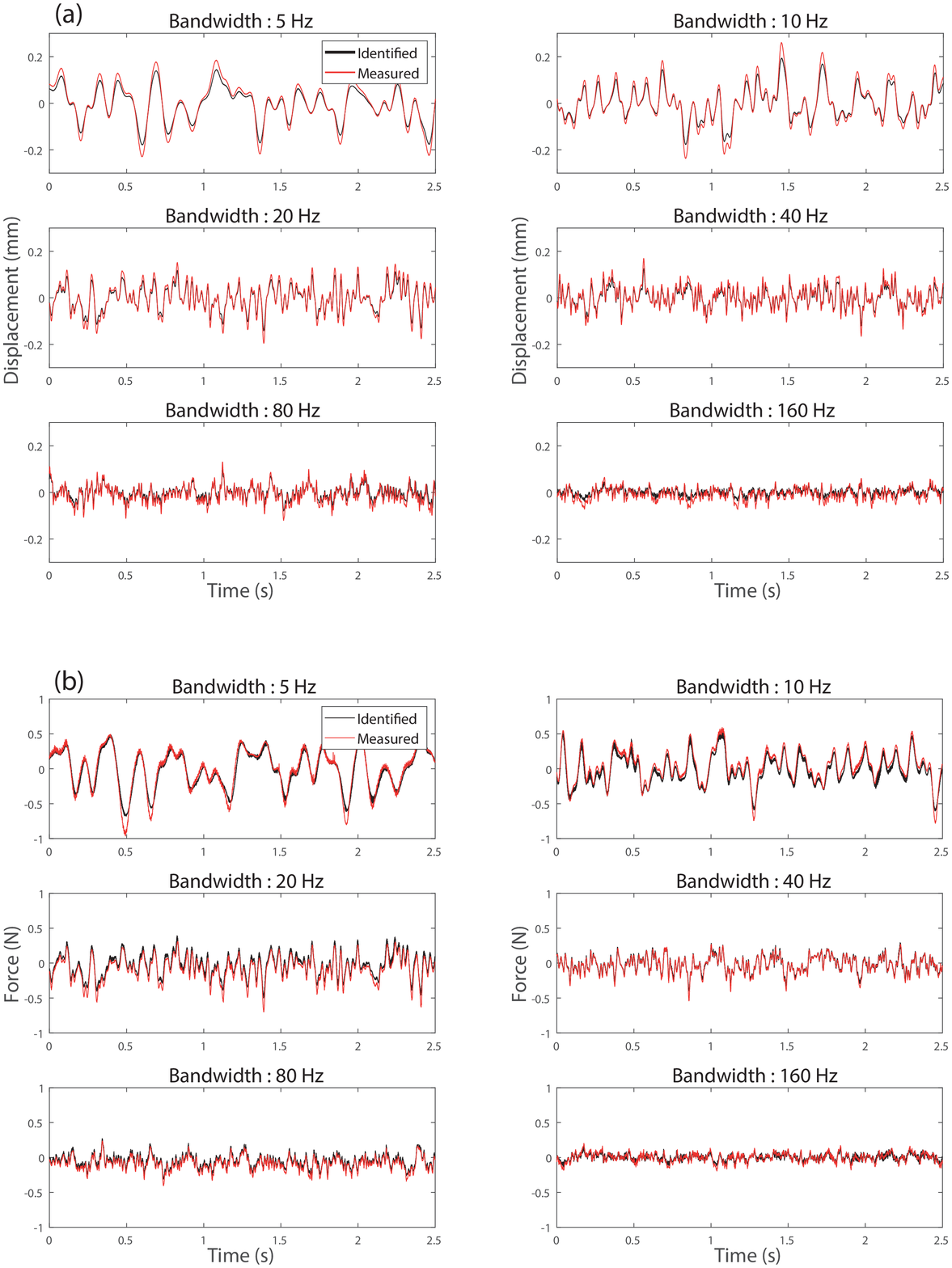}
        \caption{(a) Identified displacement results and (b) Identified force results of random excitation cases.}
        \label{fig:Exp_results}
\end{figure}

Fig.~\ref{fig:Exp_results}(a) shows the displacement identification results for the random excitation cases. The identified displacements were close to the measured displacement results.
The applied force results for random excitation cases (5-160 $\si{\Hz}$) are shown in Fig.~\ref{fig:Exp_results}(b). These results demonstrate that the system can reasonably identify unmeasured external forces.
These results reveal that the stability of the developed system can be adequately preserved under noisy conditions, and its identification results follow a similar tendency as real values, even under highly fluctuating conditions.
Table ~\ref{table:FDE_exp_random} shows the FDE index values of the identified quantities under random excitation cases.

\begin{table}[h!]
    \centering
    \caption{FDE index values of random excitation cases}
 \begin{tabular}{c|cccccc}

\hline
 Excitation frequency band & 5 $\si{\Hz}$ & 10 $\si{\Hz}$ & 20 $\si{\Hz}$ & 40 $\si{\Hz}$ & 80 $\si{\Hz}$ & 160 $\si{\Hz}$\\
 \hline
Identified force & 0.2925 & 0.1374 & 0.1690 & 0.1363 & 0.1695 & 0.1336 \\
Identified displacement & 0.1261 & 0.1374 & 0.1353 & 0.1660 & 0.1475 & 0.3275\\
\hline
\end{tabular}
    \label{table:FDE_exp_random}
\end{table}

\begin{figure}[h!]
    \centering
    \includegraphics[width=0.5\textwidth]{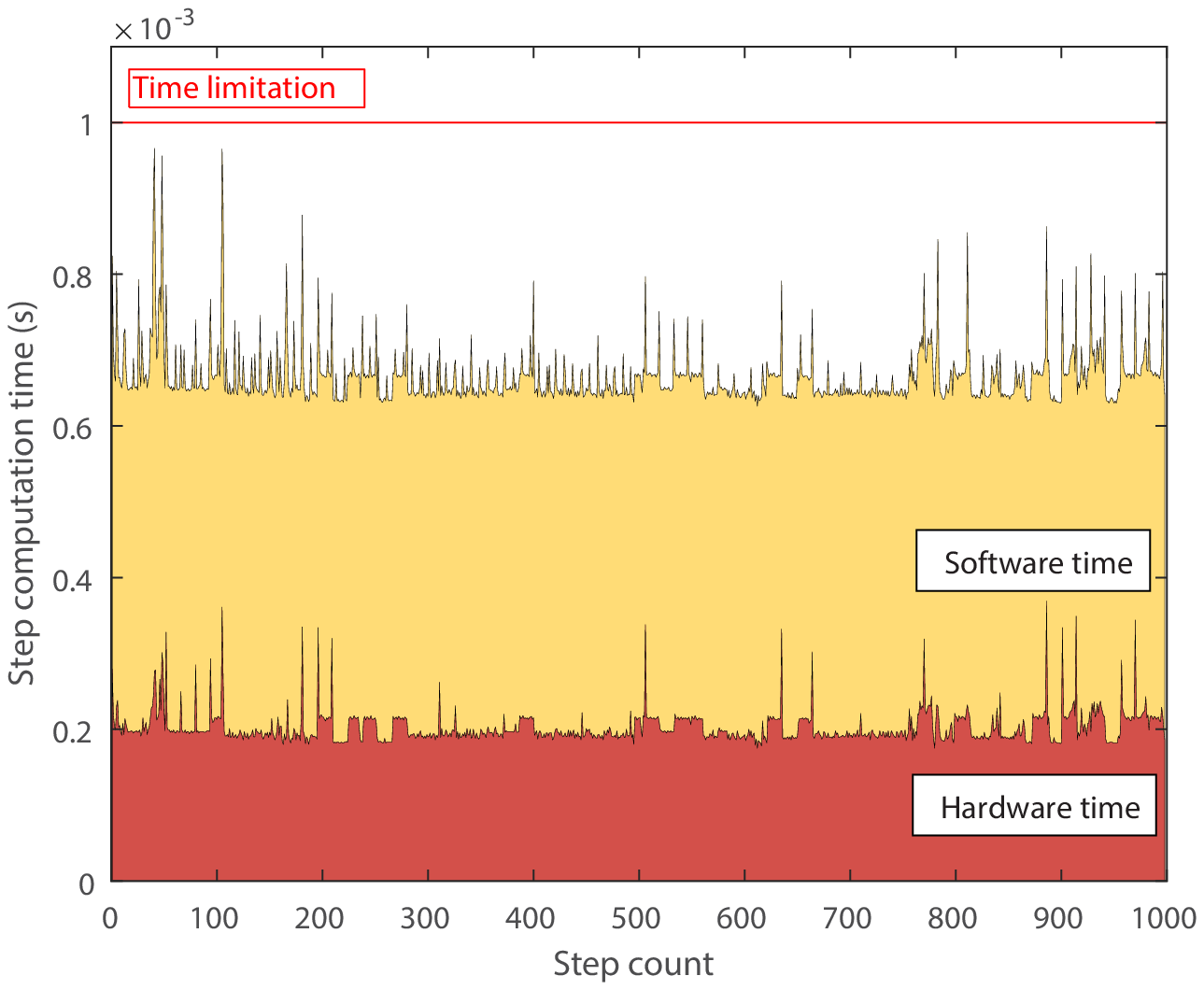}
    \caption{Computation time of every step.}
    \label{fig:Elapsedtime}
\end{figure}

The operation time of the developed virtual sensing system was evaluated to ensure stable real-time operation in an extremely limited computational environment. The operation time of the developed system can be divided into two parts. The first part is the hardware operation time, and the second is the software computation time. Hardware operation time is spent on data communication, including the analog-to-digital conversion process. The hardware operation time is highly dependent on the hardware specifications and is not involved in the real-time performance of the virtual sensing process; therefore, it is not considered here. The software computation time is the time spent on numerical calculations in the SBC using the measured data.
The required sampling rate of this system is 1,000 $Samples/s$; thus, the computation time for a single time step should be shorter than 1E-03 $s$.
The virtual sensing process is executed for 1 $s$, and the real test duration and numerical calculation time are measured for every time step.
Fig.~\ref{fig:Elapsedtime} shows the elapsed time for each iteration, and Table~\ref{table:Sinu_Random} shows the entire calculation time and the average for a single step.

% ----------------------------------------
%
% Table 7. Computation time
%
% ----------------------------------------
        
\begin{table}[h!]
    \caption{Computation time of the virtual sensing system in the example structure}
    \begin{center}
    \begin{tabular}{r r r}
    \hline
        Experimental time    & Calculation time & Single step time     \\ \hline
        1 $s$              &  0.4601 $s$    &  0.4615E-03 $s$ (Avg.)        \\ \hline

    \end{tabular}\label{table:Sinu_Random}
    \end{center}
\end{table}

The entire software calculation time required to identify the measured data response of 1 $s$ is 0.4601 $s$, and the average time required to calculate a single step is 0.4615E-3 $s$. The computation time for every iteration step was shorter than that for 1E-3 $s$. This result clearly shows that the real-time performance of the designed system satisfied the required specifications.

% ----------------------------------------
%
% Figure14. Computation time of every steps
%
% ----------------------------------------

This study tested the virtual sensing framework in a simple cantilever beam problem within 0 to 160 $\si{\Hz}$ loading conditions.
However, it could be directly extended to the higher frequency range by considering an additional precise calibration process and increasing  the counts of the dominant modes used for the ROM process.
This depends on the user's choice and certain target problems.

% ----------------------------------------
%
% 6. Conclusion
%
% ----------------------------------------
\section{Conclusions}\label{section6}
A virtual sensing algorithm of structural vibration for real-time force identification and unmeasured response prediction was proposed. Its accuracy, efficiency, and stability performance was successfully evaluated using numerical and experimental tests.
The proposed algorithm instantly and accurately identifies the response of the global structure with very limited sensors and computational resources.
To achieve real-time identification with limited resources, the following elements were mainly considered: $(\lowercase\expandafter{\romannumeral1})$  a new implicit inverse force identification method with recent advanced ROM formulation, $(\lowercase\expandafter{\romannumeral2})$  using second-order differential equations to seek better computational efficiency than the state-space form-based works, and $(\lowercase\expandafter{\romannumeral3})$ applying Tikhonov regularization for measurement noise cancelling.   
The numerical results show that the proposed algorithm accurately predicts the unmeasured forces and responses in the real-time computation process under various signal noise levels. The proposed method with $dt$ = 1E-04 provides an identification accuracy level similar to that of the conventional AKF method with $dt$ = 1E-07.  
In particular, the performance of the proposed algorithm was demonstrated in an experimental testbed. 
The prediction of the unmeasured response in the proposed method showed good agreement under sinusoidal and random excitation loading conditions, and the prediction time at each time step (including the hardware time) satisfied the required computational performance. 
The proposed virtual sensor system, a tailor-made and stand-alone sensor with low-cost hardware, has great potential for monitoring and control in vibration systems.

For future works, various advanced model updates with user-defined parameters and ROM techniques could be employed to accelerate the performance of the proposed virtual sensing algorithm instead of the methods (Newmark-$\beta$, CMS-based ROM, Tikhonov regularization, etc.) considered in this work. 
In addition, the proposed algorithm developed for time-domain analysis could be extended to frequency-domain approaches with various transfer pass analyses and system identification processes.

%\backmatter

\section*{Acknowledgments}
The research described in this paper was supported by a grant from the National Research Foundation of Korea (NRF-2020M2C9A1062790 and NRF-2021R1A2C4087079).

\section*{Author Contributions}
\textbf{Seungin Oh:} conceptualization, methodology, formal analysis, software, experiment, and writing-original draft.
\textbf{Hanmin Lee:} data curation, resources, and writing-review \& editing.
\textbf{Jaekyung Lee:} experiment support and writing-review \& editing.
\textbf{Hyungchul Yoon:} supervision, resources, and writing-review \& editing.
\textbf{Jin-Gyun Kim:} supervision, resources, investigation, funding acquisition, project administration, and writing-review \& editing.

\section*{Data Availability Statement}
The data that support the findings of this study are available from the corresponding author upon reasonable request.

\section*{ORCID}
\textit{Seungin Oh} \orcidlink{https://orcid.org/0000-0002-3213-7840}
\url{https://orcid.org/0000-0002-3213-7840}\\
\textit{Jin-Gyun Kim} \orcidlink{https://orcid.org/0000-0002-2146-7609}
\url{https://orcid.org/0000-0002-2146-7609}\\

\subsection*{Conflict of interest}

The authors declare no potential conflicts of interest.


\begin{thebibliography}{99}
\providecommand \doibase [0]{http://dx.doi.org/}%

\bibitem{doebling1996statistical}
Doebling SW, Farrar CR, Cornwell P. A statistical comparison of impact and
  ambient testing results from the Alamosa Canyon Bridge. Technical Report, Los
  Alamos National Lab, NM (United States); 1996.

\bibitem{todd1999civil}
Todd M, Johnson G, Vohra S, Chen-Chang C, Danver B, Malsawma L. Civil
  infrastructure monitoring with fiber Bragg grating sensor arrays. {\it
  Structural Health Monitoring} 1999\string: 359--368.

\bibitem{chen2015modal}
Chen JG, Wadhwa N, Cha YJ, Durand F, Freeman WT, Buyukozturk O. Modal
  identification of simple structures with high-speed video using motion
  magnification. {\it Journal of Sound and Vibration} 2015\string; 345\string:
  58--71.

\bibitem{feng2016vision}
Feng D, Feng MQ. Vision-based multipoint displacement measurement for
  structural health monitoring. {\it Structural Control and Health Monitoring}
  2016\string; 23(5)\string: 876--890.

\bibitem{yoon2016target}
Yoon H, Elanwar H, Choi H, Golparvar-Fard M, Spencer~Jr BF. Target-free
  approach for vision-based structural system identification using
  consumer-grade cameras. {\it Structural Control and Health Monitoring}
  2016\string; 23(12)\string: 1405--1416.

\bibitem{Moreau2008}
Moreau D, Cazzolato B, Zander A, Petersen C. {A review of virtual sensing
  algorithms for active noise control}. {\it Algorithms} 2008\string;
  1(2)\string: 69--99.

\bibitem{Kano2012}
Kano M, Fujiwara K. {Virtual sensing technology in process industries: trends
  and challenges revealed by recent industrial applications}. {\it Journal of
  Chemical Engineering of Japan} 2012\string: 12we167.

\bibitem{lei2020integration}
Lei Y, Lu J, Huang J. Integration of identification and vibration control of
  time-varying structures subject to unknown seismic ground excitation. {\it
  Journal of Vibration and Control} 2020\string; 26(15-16)\string: 1330--1344.

\bibitem{Ahn2022}
Ahn CU, Oh S, Kim HS, Park DI, Kim, JG. Virtual thermal sensor for real-time monitoring of electronic packages in a totally enclosed system. {\it
IEEE Access} 2022\string; 10 \string: 50589--50600.


\bibitem{Liu2009}
Liu L, Kuo SM, Zhou M. {Virtual sensing techniques and their applications}. In:
  International Conference on Networking, 2009\string: 31--36.

\bibitem{Lesniak2009}
Le{\'{s}}niak A, Danek T, Wojdy{\l}a M. {Application of Kalman filter to noise
  reduction in multichannel data}. {\it Schedae Informaticae} 2009\string;
  17(18)\string: 63--73.

\bibitem{Atkinson1998}
Atkinson CM, Long TW, Hanzevack EL. {Virtual sensing: a neural network-based
  intelligent performance and emissions prediction system for on-board
  diagnostics and engine control}. Technical Report, SAE Technical Paper; 1998.

\bibitem{Yan2016}
Yan W, Tang D, Lin Y. {A data-driven soft sensor modeling method based on deep
  learning and its application}. {\it IEEE Transactions on Industrial
  Electronics} 2016\string; 64(5)\string: 4237--4245.

\bibitem{Jain2007}
Jain P, Rahman I, Kulkarni BD. {Development of a soft sensor for a batch
  distillation column using support vector regression techniques}. {\it
  Chemical Engineering Research and Design} 2007\string; 85(2)\string:
  283--287.

\bibitem{Sun2017}
Sun SB, He YY, Zhou SD, Yue ZJ. {A data-driven response virtual sensor
  technique with partial vibration measurements using convolutional neural
  network}. {\it Sensors} 2017\string; 17(12)\string: 2888.

\bibitem{lai2021structural}
Lai Z, Mylonas C, Nagarajaiah S, Chatzi E. Structural identification with
  physics-informed neural ordinary differential equations. {\it Journal of
  Sound and Vibration} 2021\string; 508\string: 116196.


\bibitem{Risaliti2019}
Risaliti E, Tamarozzi T, Vermaut M, Cornelis B, Desmet W. {Multibody model
  based estimation of multiple loads and strain field on a vehicle suspension
  system}. {\it Mechanical Systems and Signal Processing} 2019\string;
  123\string: 1--25.

\bibitem{Kullaa2016}
Kullaa J. {Virtual sensing of structural vibrations using dynamic
  substructuring}. {\it Mechanical Systems and Signal Processing} 2016\string;
  79\string: 203--224.


\bibitem{Oh2020}
Oh S, Park D, Baek H, Kim S, Lee JK, Kim JG. {Virtual Sensing System of
  Structural Vibration using Digital Twin.}. {\it Transactions of the Korean
  Society for Noise and Vibration Engineering} 2020\string; 30(2)\string:
  149--160.
  
  \bibitem{lau1997}
  Law SS, Chan TH, Zeng QH. Moving force identification: a time domain method. {\it Journal of Sound and Vibration} 1197\string; 201(1)\string: 1-22.
  
  \bibitem{Zhu2000}
  Zhu XQ, Law SS. Identification of vehicle axle loads from bridge dynamic responses. {\it Journal of Sound and Vibration} 2000\string; 236(4) \string: 705-724.
  
  \bibitem{Zhu2001}
  Zhu XQ, Law SS. Identification of moving loads on an orthotropic plate. {\it Journal of Vibration and Acoustics} 2001\string; 123(2)\string: 238-244.
  
  \bibitem{Law2001}
  Law SS, Fang YL. Moving force identification: optimal state estimation approach. {\it Journal of Sound and Vibration} 2001\string; 239(2)\string: 233-254.
  
  \bibitem{Law2005}
  Law SS, Bu, JQ, Zhu, XQ. Time-varying wind load identification from structural responses. {\it Engineering Structures} 2005\string; 27(10)\string: 1586-1598.
  
  \bibitem{kammer1998input}
Kammer DC. Input force reconstruction using a time domain technique. 1998.

\bibitem{lourens2012augmented}
Lourens E, Reynders E, De~Roeck G, Degrande G, Lombaert G. An augmented Kalman
  filter for force identification in structural dynamics. {\it Mechanical
  systems and signal processing} 2012\string; 27\string: 446--460.
  

\bibitem{azam2015dual}
Azam SE, Chatzi E, Papadimitriou C. A dual Kalman filter approach for
  state estimation via output-only acceleration measurements, {\it Mechanical Systems and Signal Processing} 2015\string; 60\string: 866--886.

\bibitem{naets2015stable}
Naets F, Cuadrado J, Desmet W. Stable force identification in structural
  dynamics using Kalman filtering and dummy-measurements, {\it Mechanical Systems
  and Signal Processing} 2015\string; 50\string: 235--248.
  
\bibitem{liu2014explicit}
Liu K, Law SS, Zhu X, Xia Y. Explicit form of an implicit method for inverse
  force identification. {\it Journal of Sound and Vibration} 2014\string;
  333(3)\string: 730--744.



\bibitem{tikhonov1963solution}
Tikhonov AN. On the solution of ill-posed problems and the method of
  regularization. In: Doklady Akademii Nauk, Russian Academy of Sciences  1963\string:
  501--504.

\bibitem{tikhonov1995numerical}
Tikhonov AN, Goncharsky A, Stepanov V, Yagola AG. {\it Numerical methods for
  the solution of ill-posed problems}. 328.
\newblock Springer Science \& Business Media,
\newblock 1995.

\bibitem{newmark1959method}
Newmark NM. A method of computation for structural dynamics. {\it Journal of
  the Engineering Mechanics Division} 1959\string; 85(3)\string: 67--94.

\bibitem{Craig1968}
{Craig Jr} RR, Bampton MCC. {Coupling of substructures for dynamic analyses}.
  {\it AIAA Journal} 1968\string; 6(7)\string: 1313--1319.

\bibitem{Kim2015(2)}
Kim JG, Lee PS. {An enhanced Craig-Bampton method}. {\it International Journal
  for Numerical Methods in Engineering} 2015\string; 103(2)\string: 79--93.

\bibitem{Kim2017}
Kim JG, Park YJ, Lee GH, Kim DN. {A general model reduction with primal
  assembly in structural dynamics}. {\it Computer Methods in Applied Mechanics
  and Engineering} 2017\string; 324\string: 1--28.

\bibitem{Guyan1965}
Guyan RJ. {Reduction of stiffness and mass matrices}. {\it AIAA Journal}
  1965\string; 3(2)\string: 380--380.


\bibitem{Go2020}
Go MS, Lim JH, Kim JG, Hwang K. {A family of Craig--Bampton methods considering
  residual mode compensation}. {\it Applied Mathematics and Computation}
  2020\string; 369\string: 124822.

\bibitem{Bathe2006}
Bathe KJ. {\it Conserving energy and momentum in nonlinear dynamics: a simple
  implicit time integration scheme}. 85.
\newblock Elsevier,
\newblock 2007.

\bibitem{hansen1998rank}
Hansen PC. {\it Rank-deficient and discrete ill-posed problems: numerical
  aspects of linear inversion}.
\newblock SIAM,
\newblock 1998.

\bibitem{Ahn2020}
Ahn JG, Yang HI, Kim JG. Multipoint constraints with Lagrange multiplier for system dynamics and its reduced-order modeling. 
{\it AIAA Journal} 2020\string; 58(1)\string: 385--401.

\bibitem{Sherman2007}
Sherman CH, Butler JL. {\it Transducers and arrays for underwater sound}. Vol. 4. Springer, 2007.

\bibitem{dragovich2009fde}
Dragovich JJ, Lepage A. FDE index for goodness-of-fit between measured and
  calculated response signals. {\it Earthquake Engineering \& Structural
  Dynamics} 2009\string; 38(15)\string: 1751--1758.

\bibitem{taher2021earthquake}
Taher SA, Li J, Fang H. Earthquake input and state estimation for buildings
  using absolute floor accelerations. {\it Earthquake Engineering \& Structural
  Dynamics} 2021\string; 50(4)\string: 1020--1042.

\bibitem{bathe2006finite}
Bathe KJ. {\it Finite element procedures}.
\newblock Klaus-Jurgen Bathe,
\newblock 2006.

\end{thebibliography}
\end{document}